\documentclass[conference]{IEEEtran}

\usepackage{nicefrac}
\usepackage{siunitx}
\usepackage{array,framed}

\usepackage{microtype}
\renewcommand{\baselinestretch}{0.97}
\renewcommand{\baselinestretch}{0.982}
\usepackage{setspace}

\usepackage{enumitem}

\usepackage{tikz}
\usepackage{amsmath}
\usepackage{listings}
\usepackage{xspace}
\usepackage{amsthm}
\theoremstyle{definition}
\newtheorem{definition}{Definition}[section]
\usepackage{hyperref}
\usepackage{bigstrut}

\usepackage{booktabs}
\usepackage{amsthm}
\usepackage{savesym}

\usepackage{changes}

\usepackage{subcaption}
\usepackage{url}
\usepackage{amsfonts}

\usepackage{amsthm}

\usepackage{todonotes}
\setuptodonotes{fancyline, inline, color=blue!30}
\definechangesauthor[name=arthur, color=red]{arthur}

\usepackage{multirow}
\usepackage[ruled,vlined]{algorithm2e}
\usepackage{algpseudocode}
\usepackage{graphicx}
\usepackage{enumitem}
\usepackage{numprint}
\usepackage[normalem]{ulem}
\usepackage{tcolorbox}
\usepackage{xurl}
\usepackage{color}

\usepackage{caption}
\usepackage{subcaption}
\usepackage{diagbox}
\usepackage{xstring}

\definecolor{lightgray}{rgb}{.9,.9,.9}
\definecolor{darkgray}{rgb}{.4,.4,.4}
\definecolor{purple}{rgb}{0.65, 0.12, 0.82}

\lstdefinelanguage{JavaScript}{
  keywords={typeof, new, true, false, catch, function, return, null, catch, switch, var, if, in, while, do, else, case, break},
  keywordstyle=\color{blue}\bfseries,
  ndkeywords={class, export, boolean, throw, implements, import, this},
  ndkeywordstyle=\color{darkgray}\bfseries,
  identifierstyle=\color{black},
  sensitive=false,
  comment=[l]{//},
  morecomment=[s]{/*}{*/},
  commentstyle=\color{purple}\ttfamily,
  stringstyle=\color{red}\ttfamily,
  morestring=[b]',
  morestring=[b]"
}

\definecolor{verylightgray}{rgb}{.97,.97,.97}

\lstdefinelanguage{Solidity}{
	keywords=[1]{anonymous, assembly, assert, balance, break, call, callcode, case, catch, class, constant, continue, constructor, contract, debugger, default, delegatecall, delete, do, else, emit, event, experimental, export, external, false, finally, for, function, gas, if, implements, import, in, indexed, instanceof, interface, internal, is, length, library, log0, log1, log2, log3, log4, memory, modifier, new, payable, pragma, private, protected, public, pure, push, require, return, returns, revert, selfdestruct, send, solidity, storage, struct, suicide, super, switch, then, this, throw, transfer, true, try, typeof, using, value, view, while, with, addmod, ecrecover, keccak256, mulmod, ripemd160, sha256, sha3}, 
	keywordstyle=[1]\color{blue}\bfseries,
	keywords=[2]{address, bool, byte, bytes, bytes1, bytes2, bytes3, bytes4, bytes5, bytes6, bytes7, bytes8, bytes9, bytes10, bytes11, bytes12, bytes13, bytes14, bytes15, bytes16, bytes17, bytes18, bytes19, bytes20, bytes21, bytes22, bytes23, bytes24, bytes25, bytes26, bytes27, bytes28, bytes29, bytes30, bytes31, bytes32, enum, int, int8, int16, int24, int32, int40, int48, int56, int64, int72, int80, int88, int96, int104, int112, int120, int128, int136, int144, int152, int160, int168, int176, int184, int192, int200, int208, int216, int224, int232, int240, int248, int256, mapping, string, uint, uint8, uint16, uint24, uint32, uint40, uint48, uint56, uint64, uint72, uint80, uint88, uint96, uint104, uint112, uint120, uint128, uint136, uint144, uint152, uint160, uint168, uint176, uint184, uint192, uint200, uint208, uint216, uint224, uint232, uint240, uint248, uint256, var, void, ether, finney, szabo, wei, days, hours, minutes, seconds, weeks, years},	
	keywordstyle=[2]\color{teal}\bfseries,
	keywords=[3]{block, blockhash, coinbase, difficulty, gaslimit, number, timestamp, msg, data, gas, sender, sig, value, now, tx, gasprice, origin},	
	keywordstyle=[3]\color{violet}\bfseries,
	identifierstyle=\color{black},
	sensitive=false,
	comment=[l]{//},
	morecomment=[s]{/*}{*/},
	commentstyle=\color{gray}\ttfamily,
	stringstyle=\color{red}\ttfamily,
	morestring=[b]',
	morestring=[b]"
}

\DeclareMathDelimiter{(}{\mathopen} {operators}{"28}{largesymbols}{"00}
\DeclareMathDelimiter{)}{\mathclose}{operators}{"29}{largesymbols}{"01}

\setitemize{noitemsep,topsep=0pt,parsep=0pt,partopsep=0pt}
\setdescription{noitemsep,topsep=0pt,parsep=0pt,partopsep=0pt}

\makeatletter
\g@addto@macro{\normalsize}{%
    \setlength{\abovedisplayskip}{5pt}
    \setlength{\abovedisplayshortskip}{5pt}
    \setlength{\belowdisplayskip}{5pt}
    \setlength{\belowdisplayshortskip}{5pt}}
\makeatother

\newcommand{\etal}{\textit{et al.}\xspace}
\npthousandsep{,}
\npdecimalsign{.}

\usepackage{tikz}

\newcommand{\point}[1]{\par\smallskip\noindent\textbf{#1:}\xspace}

\newcommand{\empirical}[1]{#1}

\newcommand{\block}[1]{\href{https://etherscan.io/block/#1}{#1}\xspace}

\newcommand{\abbrEtherscanTx}[1]{\href{https://etherscan.io/tx/#1}{\StrLeft{#1}{6}..\StrRight{#1}{4}}\xspace}
\newcommand{\etherscanAddress}[1]{\href{https://etherscan.io/address/#1}{#1}\xspace}
\newcommand{\abbrEtherscanAddress}[1]{\href{https://etherscan.io/address/#1}{\StrLeft{#1}{6}..\StrRight{#1}{4}}\xspace}

\newcommand*\wrapletters[1]{\wr@pletters#1\@nil}
\def\wr@pletters#1#2\@nil{#1\allowbreak\if&#2&\else\wr@pletters#2\@nil\fi}

\hypersetup{
    colorlinks=true,
    linkcolor=blue,
    filecolor=magenta,
    urlcolor=cyan,
    pdfpagemode=FullScreen,
}

\usepackage{pdflscape}
\usepackage{rotating}
\usepackage{array}
\newcolumntype{L}{>{\centering\arraybackslash}m{3cm}}

\newcommand{\StartBlock}{\empirical{\block{6803256}}\xspace} 
\newcommand{\StartDate}{\empirical{1st of December,~2018}\xspace}
\newcommand{\EndBlock}{\empirical{\block{12965000}}\xspace} 
\newcommand{\EndDate}{\empirical{5th of August,~2021}\xspace}
\newcommand{\TimeDuration}{\empirical{$32$~months}\xspace}

\newcommand{\RealtimeReplayStartBlock}{\empirical{\block{12926988}}\xspace}
\newcommand{\RealtimeReplayStartDate}{\empirical{30th of July,~2021}\xspace}

\newcommand{\TotalMarkets}{\empirical{$\numprint{60830}$}\xspace}
\newcommand{\TotalTokens}{\empirical{$\numprint{49691}$}\xspace}

\newcommand{\Platforms}{\empirical{Uniswap V1/V2/V3, Sushiswap, Curve, Swerve, 1inch, and Bancor}\xspace}

\newcommand{\PlatformsSandwich}{\empirical{Uniswap V1/V2/V3, Sushiswap, and Bancor}\xspace}

\newcommand{\SandwichExtractedNumEOA}{\empirical{$\numprint{2419}$}\xspace}
\newcommand{\SandwichExtractedNumSC}{\empirical{$\numprint{1069}$}\xspace}
\newcommand{\SandwichExtractedNumAttacks}{\empirical{$\numprint{750529}$}\xspace}

\newcommand{\SandwichExtractedTotalProfit}{\empirical{$\numprint{174.34}$M~USD}\xspace}
\newcommand{\SandwichExtractedNotSameAddress}{\empirical{$17.57\%$}\xspace}

\newcommand{\SandwichExtractedHFTPaper}{\empirical{$62.13$~ETH}\xspace}
\newcommand{\SandwichExtractedPercentage}{\empirical{$63.30\%$}\xspace}
\newcommand{\SandwichExtractedPerfect}{\empirical{$\numprint{603431}$}\xspace}
\newcommand{\SandwichExtractedPerfectPercentage}{\empirical{$80.4\%$}\xspace}

\newcommand{\SandwichExtractedZeroGasPrice}{\empirical{$\numprint{240053}$}\xspace}
\newcommand{\SandwichExtractedPercentageZeroGasPrice}{\empirical{$31.98$\%}\xspace}
\newcommand{\SandwichExtractedZeroGasPriceProfit}{\empirical{$\numprint{81.04}$M~USD}\xspace}

\newcommand{\SandwichExtractedImperfect}{\empirical{$\numprint{147098}$}\xspace}
\newcommand{\SandwichExtractedImperfectPercentage}{\empirical{$19.6\%$}\xspace}

\newcommand{\SandwichPrivatelyRelayedPercentageCLose}{\empirical{$99.59\%$}\xspace}

\newcommand{\SandwichPercentageBumpFive}{\empirical{$19.11\%$}\xspace}
\newcommand{\SandwichPercentageOneWei}{\empirical{$80.02\%$}\xspace}

\newcommand{\LiquidationPlatforms}{\empirical{Aave V1/V2, Compound, and dYdX}\xspace}

\newcommand{\TotalNumFixedSpreadLiquidation}{\empirical{$\numprint{31057}$}\xspace}
\newcommand{\ZeroGasPriceLiquidation}{\empirical{$\numprint{1956}$}\xspace}
\newcommand{\ZeroGasPriceLiquidationPercentage}{\empirical{$6.3\%$}\xspace}
\newcommand{\ZeroGasPriceLiquidationProfit}{\empirical{$10.69$M~USD}\xspace}
\newcommand{\LiquidationCrawlingPeriod}{\empirical{$28$~months}\xspace}
\newcommand{\AaveVTwoLiquidations}{\empirical{$\numprint{6432}$}\xspace}
\newcommand{\AaveVOneLiquidations}{\empirical{$\numprint{4932}$}\xspace}
\newcommand{\CompoundLiquidations}{\empirical{$\numprint{9287}$}\xspace}
\newcommand{\dYdXLiquidations}{\empirical{$\numprint{10406}$}\xspace}
\newcommand{\TotalFixedSpreadLiquidationProfit}{\empirical{$89.18$M~USD}\xspace}

\newcommand{\AaveVTwoFrontRunning}{\empirical{$\numprint{4085}$}\xspace}
\newcommand{\AaveVTwoBackRunning}{\empirical{$\numprint{2347}$}\xspace}
\newcommand{\AaveVOneFrontRunning}{\empirical{$\numprint{4331}$}\xspace}
\newcommand{\AaveVOneBackRunning}{\empirical{$\numprint{601}$}\xspace}
\newcommand{\CompoundFrontRunning}{\empirical{$\numprint{6119}$}\xspace}
\newcommand{\CompoundBackRunning}{\empirical{$\numprint{3168}$}\xspace}
\newcommand{\dYdXFrontRunning}{\empirical{$\numprint{8603}$}\xspace}
\newcommand{\dYdXBackRunning}{\empirical{$\numprint{1803}$}\xspace}
\newcommand{\TotalLiquidationFrontRunning}{\empirical{$\numprint{23138}$}\xspace}
\newcommand{\TotalLiquidationBackRunning}{\empirical{$\numprint{7919}$}\xspace}
\newcommand{\LiquidationFrontRunningPercentage}{\empirical{$74.50\%$}\xspace}

\newcommand{\TotalLiquidators}{\empirical{$\numprint{2742}$}\xspace}
\newcommand{\FrontRunningLiquidators}{\empirical{$\numprint{1758}$}\xspace}
\newcommand{\BackRunningLiquidators}{\empirical{$\numprint{442}$}\xspace}
\newcommand{\MixedLiquidators}{\empirical{$\numprint{542}$}\xspace}

\newcommand{\ArbitrageExtractedNumEOA}{\empirical{$\numprint{6753}$}\xspace}
\newcommand{\ArbitrageExtractedNumSC}{\empirical{$\numprint{2016}$}\xspace}
\newcommand{\ArbitrageExtractedNumAttacks}{\empirical{$\numprint{1151448}$}\xspace}
\newcommand{\ArbitrageExtractedTotalProfit}{\empirical{$277.02$M~USD}\xspace}
\newcommand{\ArbitrageBackRunning}{\empirical{$44.02\%$}\xspace}

\newcommand{\ZeroGasPriceArbitrage}{\empirical{$\numprint{110026}$}\xspace}
\newcommand{\ZeroGasPriceArbitragePercentage}{\empirical{$9.6\%$}\xspace}
\newcommand{\ZeroGasPriceArbitrageProfit}{\empirical{$82.75$M~USD}\xspace}

\newcommand{\PositiveGasPriceArbitrage}{\empirical{$\numprint{1041422}$}\xspace}

\newcommand{\CloggingNumEOA}{\empirical{$10$}\xspace}
\newcommand{\CloggingNumSC}{\empirical{$75$}\xspace}
\newcommand{\CloggingNumAttacks}{\empirical{$\numprint{333}$}\xspace}

\newcommand{\ReplayExtendBEV}{\empirical{$\numprint{35.18}$M~USD}\xspace}

\newcommand{\ReplayPerformance}{\empirical{$0.18\pm0.29$~seconds}\xspace}

\newcommand{\TotalNaiveReplayTestTransactions}{\empirical{$\numprint{883023232}$}\xspace}

\newcommand{\ReplayableTransctionInJuneTwentyTwenty}{\empirical{$\numprint{19825}$}\xspace}

\newcommand{\TotalReplayableTransactions}{\empirical{$\numprint{188365}$}\xspace}
\newcommand{\TotalReplayableTransactionPercentage}{\empirical{$0.02\%$}\xspace}

\newcommand{\TotalReplayProfit}{\empirical{$\numprint{57037.32}$~ETH}\xspace}
\newcommand{\TotalReplayProfitUSD}{\empirical{$35.37$M~USD}\xspace}

\newcommand{\MostProfitableReplay}{\empirical{$\numprint{16736.9}$~ETH}\xspace}
\newcommand{\BeneficiarysenderTransactions}{\empirical{$\numprint{171219}$}\xspace}
\newcommand{\SpecifyBeneficiaryTransactions}{\empirical{$\numprint{17146}$}\xspace}

\newcommand{\ReplayTransactionTokenNumber}{\empirical{$\numprint{1213}$}\xspace}
\newcommand{\ReplayTransactionTokenRevenue}{\empirical{$\numprint{179843.52}$~ETH}\xspace}
\newcommand{\ReplayWithTokenTransactions}{\empirical{$\numprint{128200}$}\xspace}
\newcommand{\TotalTransactionsYieldsOverOneETH}{\empirical{$\numprint{1926}$}\xspace}
\newcommand{\TotalZeroTransactionsYieldsOverOneETH}{\empirical{$\numprint{1007}$}\xspace}

\newcommand{\ReplayableLiquidationTransactions}{\empirical{$443$}\xspace}
\newcommand{\ReplayableLiquidationProfit}{\empirical{$20.44$K~USD}\xspace}
\newcommand{\ReplayableArbitrageTransactions}{\empirical{$\numprint{1268}$}\xspace}
\newcommand{\ReplayableArbitrageProfit}{\empirical{$165.38$K~USD}\xspace}

\newcommand{\ZeroGasPriceReplayableTransaction}{\empirical{$\numprint{6685}$}\xspace}
\newcommand{\ZeroGasPriceReplayableTransactionProfit}{\empirical{$\numprint{3.63}$M~USD}\xspace}

\newcommand{\OverFourTimesBEVBlockNum}{\empirical{$\numprint{2407}$}\xspace}
\newcommand{\TotalExtractedBEV}{\empirical{$540.54$M~USD}\xspace}

\newcommand{\MaxSingleBlockBEVETH}{\empirical{$8,453.9$~ETH}\xspace}

\newcommand{\MaxSingleBlockBEVUSD}{\empirical{$4.1$M~USD}\xspace}
\newcommand{\MaxSingleBlockBlockNum}{\empirical{\block{11333037}}}
\newcommand{\MaxSingleBlockBEVMultiple}{\empirical{$616.6$}\xspace}
\newcommand{\MaxSingleBlockBEVMultipleTimes}{\empirical{$616.6\times$}\xspace}

\begin{document}

\title{Quantifying Blockchain Extractable Value: \\How dark is the forest?}

\author{\IEEEauthorblockN{Kaihua Qin}
\IEEEauthorblockA{Imperial College London\\
kaihua.qin@imperial.ac.uk}
\and
\IEEEauthorblockN{Liyi Zhou}
\IEEEauthorblockA{Imperial College London\\
liyi.zhou@imperial.ac.uk}
\and
\IEEEauthorblockN{Arthur Gervais}
\IEEEauthorblockA{Imperial College London\\
a.gervais@imperial.ac.uk}}

\maketitle
\thispagestyle{plain}
\pagestyle{plain}

\begin{abstract}
Permissionless blockchains such as Bitcoin have excelled at financial services. Yet, opportunistic traders extract monetary value from the mesh of decentralized finance (DeFi) smart contracts through so-called blockchain extractable value (BEV). The recent emergence of centralized BEV relayer portrays BEV as a positive additional revenue source. Because BEV was quantitatively shown to deteriorate the blockchain's consensus security, BEV relayers endanger the ledger security by incentivizing rational miners to fork the chain. For example, a rational miner with a $10$\% hashrate will fork Ethereum if a BEV opportunity exceeds $4\times$ the block reward.

However, related work is currently missing quantitative insights on past BEV extraction to assess the practical risks of BEV objectively. In this work, we allow to quantify the BEV danger by deriving the USD extracted from sandwich attacks, liquidations, and decentralized exchange arbitrage. We estimate that over~\TimeDuration, BEV yielded~\TotalExtractedBEV in profit, divided among~\empirical{$\numprint{11289}$} addresses when capturing~\TotalTokens cryptocurrencies and~\TotalMarkets on-chain markets. The highest BEV instance we find amounts to \MaxSingleBlockBEVUSD, \MaxSingleBlockBEVMultipleTimes the Ethereum block reward.

Moreover, while the practitioner's community has discussed the existence of generalized trading bots, we are, to our knowledge, the first to provide a concrete algorithm. Our algorithm can replace unconfirmed transactions without the need to understand the victim transactions' underlying logic, which we estimate to have yielded a profit of \TotalReplayProfit (\TotalReplayProfitUSD) over~\TimeDuration of past blockchain data.

Finally, we formalize and analyze emerging BEV relay systems, where miners accept BEV transactions from a centralized relay server instead of the peer-to-peer (P2P) network. We find that such relay systems aggravate the consensus layer attacks and therefore further endanger blockchain security.

\end{abstract}

\section{Introduction}\label{sec:intro}
With a locked value of over $90$B USD in Decentralized Finance (DeFi), distributed ledgers have shown their strength in mediating trustlessly among financial actors exchanging daily hundreds of millions of USD. DeFi traders rely on immutable smart contracts encoding the rules by which, for instance, automated market maker (AMM) exchanges~\cite{httpsuni34:online} operate. DeFi on permissionless blockchains operates surprisingly transparent compared to the traditional finance. All transactions, sender, receiver and amounts are publicly visible on a global P2P network, prior to being committed by miners to the ledger. Miners herein retain the privilege to control single-handedly the transaction order of their mined blocks, an information asymmetry which is being exploited for financial gain~\cite{zhou2021high}.

Besides miners, blockchain value extracting traders have specialized in maximizing financial revenue through ongoing market participation. Similar to the traditional finance, DeFi is being plagued by predatory traders, showcasing a plethora of creative market manipulation techniques, such as high-frequency attacks~\cite{zhou2021high}, pump and dump schemes~\cite{xu2019anatomy} and wash trading~\cite{victor2021detecting}. Akin to how Eskandir \etal~\cite{eskandari2019sok} beautifully distill the state of open and decentralized ledgers: we observe a distributed network of transparent dishonesty --- once a user broadcasts a profitable transaction, seemingly automated trading-bots attempt to appropriate the trading opportunity by front-running their victim with higher transaction fees~\cite{darkforest} to extract \emph{blockchain extractable value} (BEV).

The existence of BEV appears to radically transform the distributed ledger incentive structure. Previous studies~\cite{daian2020flash,zhou2021just} suggest, and show, that miners are incentivized to extract value by deliberately forking a chain, endangering blockchain security. To the best of our knowledge, no work has yet comprehensively measured and studied the real-world severity of BEV. Quantifying the status quo of BEV, however, is crucial to understand the risks that blockchain users are exposed to.

\begin{figure}[bt]
    \centering
    \includegraphics[width=\columnwidth]{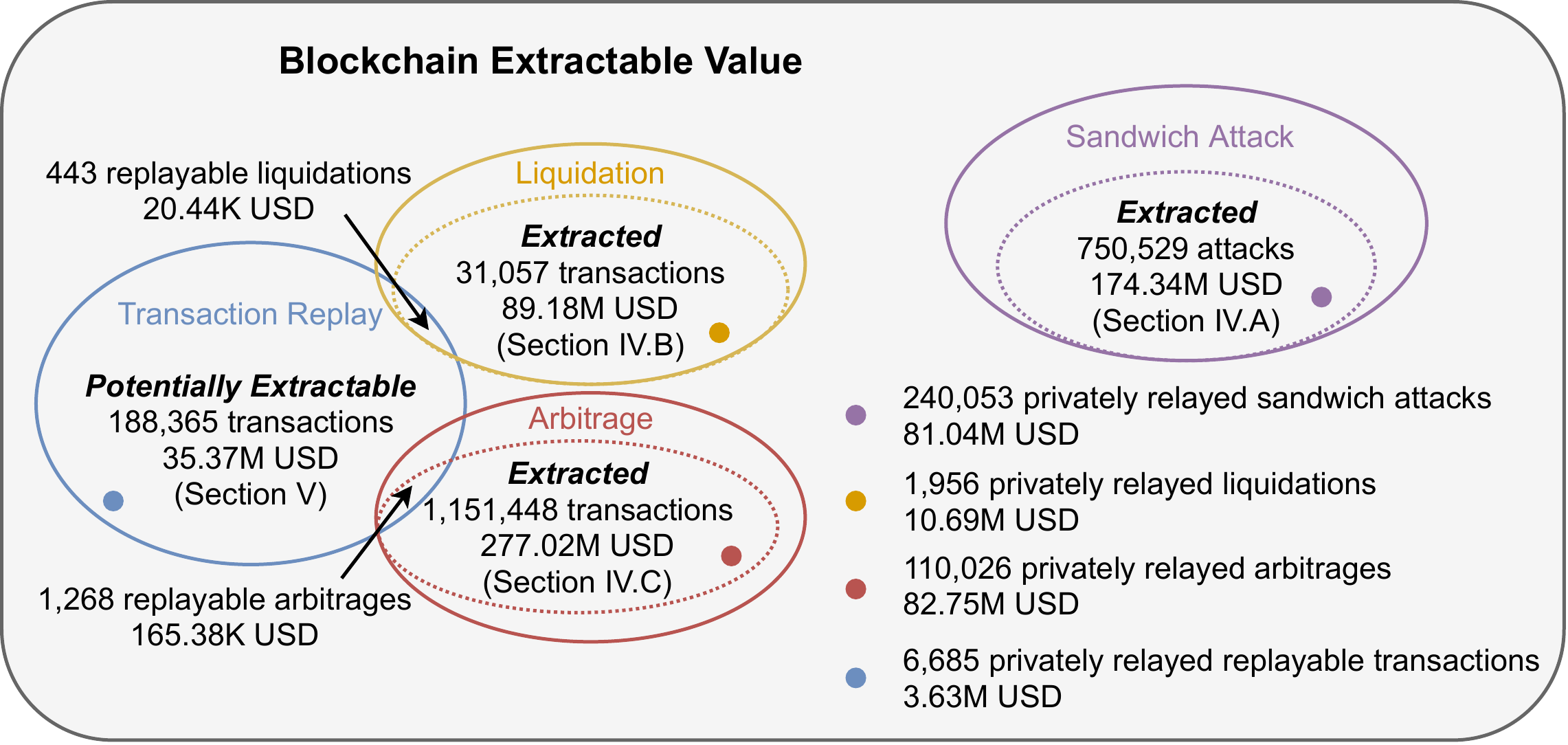}
    \caption{Overview of various sources of blockchain extractable value. We find that sandwich attacks, liquidations and arbitrage yield~\TotalExtractedBEV of BEV over~\TimeDuration. We further evaluate a novel application-agnostic transaction replay algorithm, which could have extended BEV by~\ReplayExtendBEV.}
    \label{fig:BEV-overview}
\end{figure}

In this work we capture a variety of BEV sources, including sandwich attacks, liquidations, and arbitrage (cf.\ Fig.~\ref{fig:BEV-overview} and Section~\ref{sec:measurement}). We moreover present the first generalized transaction replay algorithm, which allows to clone and front-run a victim transaction without the need to understand the underlying victim transaction logic (cf.\ Section~\ref{sec:naivereplay}). The potential extractable value from transaction replay attacks can significantly extend the total BEV (cf.\ Fig.~\ref{fig:BEV-overview}), further endangering the blockchain security.

More worryingly, we observe the recent emergence of centralized BEV relayer (e.g., flashbots). A BEV relayer acts as a proxy between BEV traders and miners, filtering the trades that are forwarded to the miners. The goal of a BEV relayer is to maximize BEV, and hence in expectation, increases the number of blockchain forks and chain reorganizations~\cite{zhou2021just}.

Summarizing, our main contributions are as follows.

\begin{itemize}
    \item We are the first to comprehensively measure the breadth of BEV from known trading activities (i.e., sandwich attacks, liquidations, and arbitrages). Although related works have studied sandwich attacks in isolation, there is a lack of quantitative data from real-world exploitation to objectively assess their severity.
    \item We are the first to propose and empirically evaluate a transaction replay algorithm, which could have resulted in \TotalReplayProfitUSD of BEV. Our algorithm extends the total captured BEV by~\ReplayExtendBEV, while intersecting with only $1.43$\% of the liquidation and $0.11$\% of the arbitrage transactions (cf.\ Fig.~\ref{fig:BEV-overview}).
    \item We are the first to formalize the BEV relay concept as an extension of the P2P transaction fee auction model. Contrary to the suggestions of the practitioner community, we find that a BEV relayer does not substantially reduce the P2P network overhead from competitive trading.
\end{itemize}

\section{Background}\label{sec:background}
\subsection{Blockchain and Smart Contracts}
Permissionless blockchains are span by a network of globally distributed P2P nodes~\cite{bitcoin}. If a user wishes to execute a transaction on the blockchain (which in essence is a distributed database), the user broadcasts the transaction to its P2P neighbors. These neighbors then forward that transaction until the transaction eventually reaches a miner. A miner constructs a block to append data to the blockchain and decides unilaterally on the transactions execution order. A transaction that is included in at least one blockchain block (i.e., the chain with most ``Proof of Work'') is considered confirmed (i.e., a one-confirmation) by the network. Blockchains differ in confirmation latencies, ranging from hours in Bitcoin~\cite{bitcoin} to minutes in Ethereum~\cite{wood2014ethereum}, while offering distinct security trade-offs~\cite{gervais2016security}. Generally, there is an inherent \emph{time delay}, between the public \emph{broadcast} of a transaction and its execution. Blockchain nodes store unconfirmed transactions within the so-called \emph{mempool}. For a more thorough background, we refer the reader to helpful SoKs~\cite{bonneau2015sok, atzei2017survey, bano2019sok}.

We proceed to outline the required background of Ethereum. Beyond simple value transfers, Etherum is a smart contract-enabled blockchain~\cite{wood2014ethereum}, which allows the construction of DeFi protocols. Smart contracts execute within a virtual machine called Ethereum Virtual Machine (EVM). In this paper, we differentiate among user addresses (i.e., owned by a private key) and smart contract addresses. In Ethereum, blocks can be indexed by the block number, an incremental integer, while transactions are often indexed by the transaction hash, the Keccak-256 hash value of a transaction. ETH is the native cryptocurrency in Ethereum, which can be used to, for example, pay transaction fees. The transaction fee is calculated with \emph{gas} (measuring the amount computations consumed in a transaction) times \emph{gas price} (the amount that the transaction issuer is willing to pay for each unit of gas). The smallest unit of ETH is Wei, equivalent to~$10^{-18}$ ETH. Transaction fees are commonly denominated in GWei (i.e., $10^9$ Wei).
In addition to a chain's native cryptocurrency, smart contracts allow to create on-chain assets, so-called tokens. At the time of writing, ERC20 is the most widely adopted token standard.

\subsection{Decentralized Finance}
DeFi is a subset of finance-focused decentralized protocols that operate autonomously on blockchain-based smart contracts~\cite{qin2021cefi}. After excluding the DeFi systems' endogenous assets, the total value locked in DeFi amounts to $90$B USD at the time of writing. Relevant DeFi platforms are for instance automated market maker exchanges~\cite{httpsuni34:online, Hertzog2018}, lending platforms~\cite{aave,dydx, makerdao, compoundfinance} and margin trading systems~\cite{bzxnetwork}.

\point{AMM Exchanges}
Traditional limit order-book-based exchanges maintain a list of bids and asks for an asset pair. AMM exchanges, however, maintain a pool of capital (i.e., a liquidity pool) with at least two assets. A smart contract governs the rules by which traders can purchase and sell assets from the liquidity pool. The most common AMM mechanism is the constant product rule in a pair-asset market. This rule stipulates that the product of an asset $x$ and asset $y$ in a pool remains a constant $k$. Uniswap, with over $8$B USD total value locked (TVL), one of the biggest AMM exchanges at the time of writing, follows a constant product AMM model~\cite{httpsuni34:online}.

\point{Slippage}
When performing a trade on an AMM, a trader is exposed to an expected slippage depending on the available liquidity in the AMM (i.e., the price gets worse as the trading volume increases). Furthermore, the expected execution price may differ from the real execution price (i.e., an unexpected slippage). That is because the expected price is derived upon a past blockchain state, which may change between the transaction creation and its execution --- e.g., due to front-running transactions~\cite{zhou2021high}. Therefore, a trader typically sets a slippage tolerance (i.e., the maximum acceptable slippage) when issuing an AMM trading transaction.

\point{Lending Systems}
Debt is an essential tool in traditional finance~\cite{dalio2012economic}, and the same applies to DeFi. 
DeFi lending typically requires over-collateralization~\cite{qin2021empirical}. Hence, a borrower must collateralize, i.e., lock, for instance, $150$\% of the value that the borrower wishes to lend out. The collateral acts as a security fund to the lender if the borrower does not pay back the debt. If the collateral value decreases and the collateralization ratio decreases below $150$\%, the collateral can be freed up for liquidation. Liquidators can then purchase the collateral at a discount to repay the debt. At the time of writing, lending systems on the Ethereum blockchain have accumulated a TVL of $40$B USD~\cite{aave,dydx,makerdao,compoundfinance}.

\section{Preliminaries}\label{sec:preliminaries}
In this section, we outline our security and threat model. We discuss how the blockchain transaction order relates to BEV and proceed with a blockchain transaction ordering taxonomy.

\subsection{System and Threat Model}
We consider a permissionless blockchain system on top of a P2P network. We assume the existence of a trader $V$ conducting at least one blockchain transaction $T_V$ (given a public/private key-pair) by, e.g., trading assets on AMM exchanges or interacting with a lending platform. The trader is free to specify its slippage tolerance, transaction fees, and choice of platform. We refer to the trader as a victim if other traders attack the trader (e.g., in a sandwich attack). We further assume the existence of a set of miners that may or may not engage in extracting blockchain extractable value. The miners can choose to order transactions according to internal policies or may follow the transaction fee distribution.

Our threat model captures a financially rational adversary $\mathcal{A}$ that is well-connected in the network layer to observe unconfirmed transactions in the memory pool. $\mathcal{A}$ holds at least one private key for a blockchain account from which it can issue an authenticated transaction $T_A$. We also assume that $\mathcal{A}$ owns a sufficient balance of the native cryptocurrency (e.g., ETH on Ethereum) to perform actions required by $T_A$, e.g., paying transaction fees or trading assets. If $\mathcal{A}$ is a mining entity, then $\mathcal{A}$ can unilaterally decide which and in which order transactions figure within its mined blocks. When $\mathcal{A}$ is a non-mining entity, $\mathcal{A}$ attempts to extract value by adjusting the transaction fees or resorting to BEV relayers (cf.\ Section~\ref{sec:bev-relayer}).

\subsection{Transaction Ordering and Blockchain Extractable Value}
Compared to traditional financial systems (e.g., centralized exchanges), we identify that the value extraction game on blockchains presents two fundamental properties.

\point{Atomicity} Multiple actions fit into one transaction and execute in an all-or-nothing sequence~\cite{allen2020design,qin2021attacking}. If a single action of an atomic transaction fails, all previously executed actions are reverted without permeating a blockchain state change.

\point{Determinacy} Given a blockchain state, the execution of a transaction is deterministic. Trader can hence simulate or ``predict'' the execution result before a transaction is mined.

These two properties are decisive for the value extraction game. An adversary attempts to manipulate the transaction order, such that the adversarial transactions execute on a blockchain state which maximizes the adversarial revenue. The order manipulation may prioritize adversarial transactions or attempt to move a victim transaction to execute on an unfavorable blockchain state. We provide a detailed transaction ordering taxonomy in Section~\ref{sec:transaction-order-taxonomy}.

Previous works have shown how trading bots engage in competitive transaction fee bidding contests~\cite{daian2020flash, zhou2021high}. Besides exchange trading, front-running was observed on blockchain games, crypto-collectibles, gambling, ICOs, and name services~\cite{eskandari2019sok}. Miner Extractable Value, first introduced by Daian \etal~\cite{daian2020flash}, captures the blockchain extractable value from miners. However, non-mining traders can also capture BEV by adjusting, for example, their transaction fees, and we observe MEV as a subset of the blockchain extractable value.

\subsection{Transaction Ordering Taxonomy}\label{sec:transaction-order-taxonomy}
\begin{figure}[t!]
    \centering
    \includegraphics[width=\columnwidth]{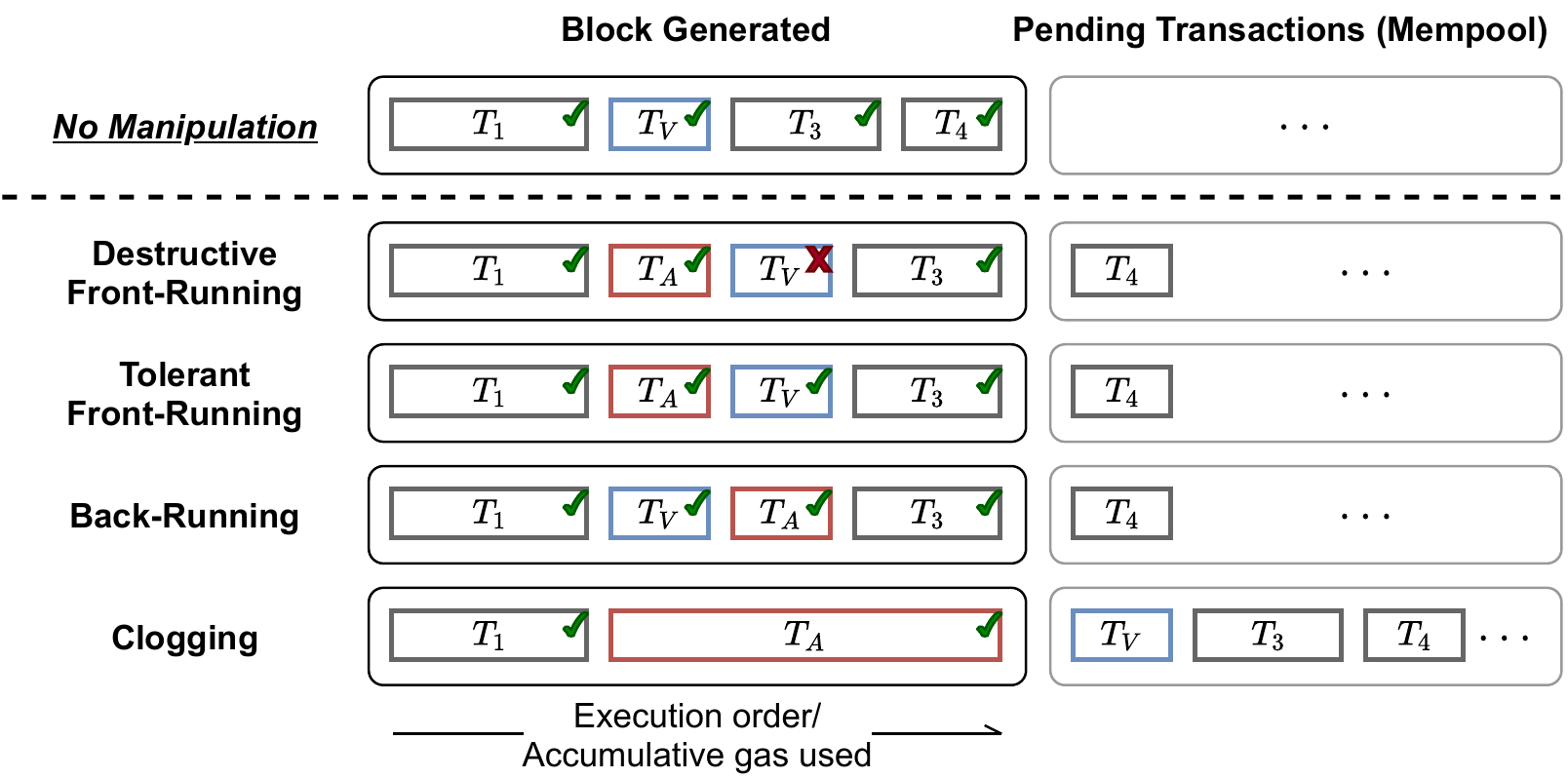}
    \caption{Visualization of four adversarial transaction ordering strategies. $T_V$ is the victim and $T_A$ the adversarial transaction. $T_1$ to $T_4$, are included in that sequence in the next block.}
    \label{fig:taxonomy}
\end{figure}

In light of the decisive pertinence of the transaction order on blockchain value extraction, we provide in the following a transaction ordering taxonomy which extends the three front-running categories discussed in related work~\cite{eskandari2019sok}. We explicitly add a fourth category, which captures the act of back-running a transaction (cf.\ Fig.~\ref{fig:taxonomy}). We moreover highlight the subtle but essential impact of an adversarial front-running transaction on the subsequent victim transaction: either $T_A$ provokes the victim transaction to fail, or the adversary takes care to avoid that $T_V$ reverts after a successful front-running.

\begin{table}[b]
\centering
\caption{Attack surface for non-mining adversaries. Sandwich attacks and transaction replay occur on the network state.}
\resizebox{0.8\columnwidth}{!}{%
\begin{tabular}{@{}lll@{}}
\toprule
\textbf{Use Case}        & \textbf{Block State}   & \textbf{Mempool/Network State} \\ \midrule
Sandwich Attack & -             & \checkmark \\
Liquidation     & \checkmark    & \checkmark (back-running oracles) \\
Arbitrage       & \checkmark    & \checkmark \\
Transaction Replay          & -             & \checkmark \\ \bottomrule
\end{tabular}%
}
\label{tab:attack-states}
\end{table}

\begin{figure*}
    \centering
    \includegraphics[width=\textwidth]{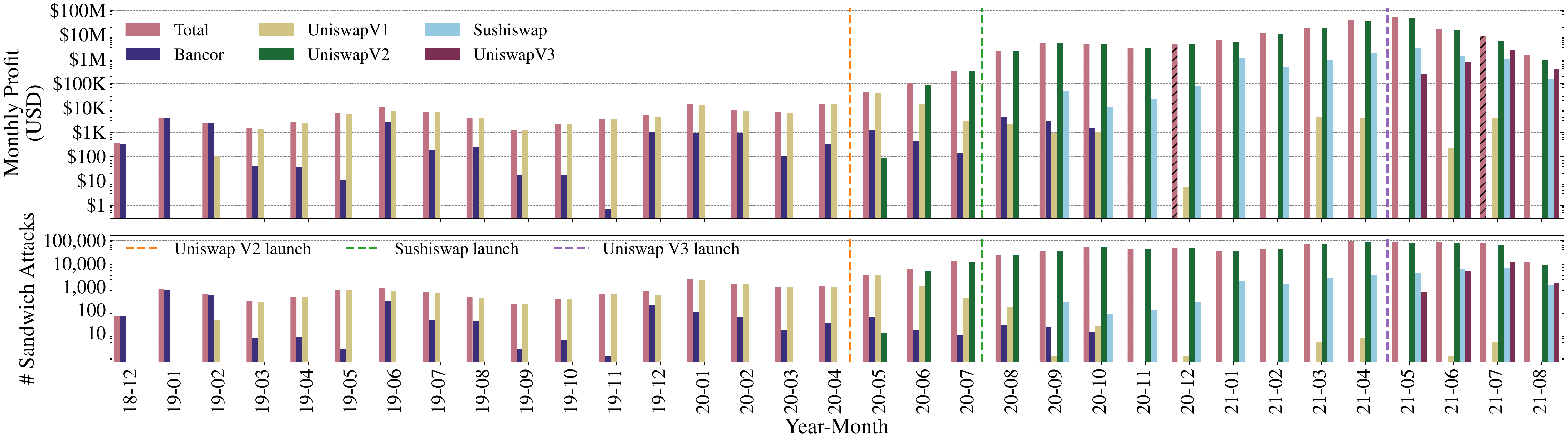}
    \caption{Sandwich attacks, from block~\StartBlock (\StartDate) to~\EndBlock (\EndDate).}
    \label{fig:sandwich-monthly-statistics}
\end{figure*}

\point{Destructive Front-Running} If $T_A$ front-runs $T_V$, and causes the execution of $T_V$ to fail (i.e., the EVM reverts the transaction state changes), we classify the act of front-running as destructive. The front-running adversary, therefore, bears no considerations about its impact on subsequent transactions.
\point{Tolerating Front-Running} Front-running is ``tolerating'', if the adversary ensures that $T_V$ executes successfully. Tolerating front-running is necessary for, e.g., sandwich attacks~\cite{zhou2021high}. An adversary would not be able to profit from sandwich attacks with destructive front-running.
\point{Back-Running} Executing $T_A$ after $T_V$ is called back-running, a technique which can be applied after, e.g., oracle update transactions~\cite{liu2020first,eskandari2021sok} and within sandwich attacks~\cite{zhou2021high}. Back-running is, in expectation, cheaper than front-running, as the trader does not engage in a fee bidding contest.
\point{Clogging} An adversary may clog, or jam the blockchain with transactions, to prevent users and bots from issuing transactions (i.e., suppression~\cite{eskandari2019sok}). Deadline-based smart contracts may create an incentive to clog the blockchain.

Transaction ordering may occur on different blockchain state representations. We differentiate in this paper between a block state and a mempool/network state (cf.\ Table~\ref{tab:attack-states}). A block state corresponds to the last confirmed main-chain head, while the mempool state is a more volatile and local state of a blockchain P2P node. We notice that sandwich attacks (cf.\ Section~\ref{sec:sandwich}) and transaction replay (cf.\ Section~\ref{sec:naivereplay}) can only occur on the network layer (unless a miner forks the blockchain).

\section{Measuring the Extracted Blockchain Value}\label{sec:measurement}
In the following, we investigate to what extent traders have extracted financial value from the Ethereum blockchain over a time frame of \TimeDuration (from the \StartDate to the \EndDate). While it is challenging to capture all possible revenue strategies, we do not claim completeness and choose to focus on sandwich attacks, liquidations, and arbitrage trading. For the sandwich and arbitrage, we inspect all the trades performed on \Platforms, spanning over~\TotalTokens cryptocurrencies and~\TotalMarkets on-chain markets. For liquidations, we collect every liquidation event settled on \LiquidationPlatforms. Throughout our measurement, we identify transactions with zero gas price as privately relayed transactions\footnote{Transactions with zero gas price are not propagating on the Ethereum P2P network due to DoS concerns. Miners, however, might receive these transactions from, for example, BEV relayers (cf.\ Section~\ref{sec:bev-relayer}).}.




\subsection{Sandwich Attacks}\label{sec:sandwich}
Sandwich attacks, wherein a trader wraps a victim transaction within two adversarial transactions, is a classic predatory trading strategy~\cite{zhou2021high}. To perform a sandwich, the adversary $\mathcal{A}$, which can be a miner or trader, listens on the P2P network for pending transactions. The adversary attacks, if the market price of an asset is expected to rise/fall after the execution of a ``large'' pending transaction ($T_V$). The attack is then carried out in two-steps: \emph{(i)} $\mathcal{A}$ issues $T_{A1}$ to \textbf{tolerating front-run} $T_V$, by purchasing/selling the same asset before $T_V$ changes the market price; \emph{(ii)} $\mathcal{A}$ then issues $T_{A2}$ to \textbf{back-run} $T_V$ to close the trading position opened by $T_{A1}$. $\mathcal{A}$ must perform tolerating front-running to ensure that $T_V$'s slippage protection does not trigger a transaction revert.

\subsubsection{Heuristics} We apply the following heuristics to identify potentially successful sandwich attacks from the AMM trades.
\begin{itemize}[left=0pt]
    \item \textbf{Heuristic 1:} The transactions $T_{A1}$, $T_{V}$ and $T_{A2}$ must be included in the same block and in this exact order.
    \item \textbf{Heuristic 2:} Every front-running transaction $T_{A1}$ maps to \textbf{one and only one} back-running transaction $T_{A2}$. This heuristic is necessary to avoid double counting revenues.
    \item \textbf{Heuristic 3:} Both $T_{A1}$ and $T_{V}$ transact from asset $X$ to $Y$. $T_{A2}$ transacts in the reverse direction from asset $Y$ to $X$.
    \item \textbf{Heuristic 4:} Either the same user address sends transactions $T_{A1}$ and $T_{A2}$, or two different user addresses send $T_{A1}$ and $T_{A2}$ to the same smart contract.
    \item \textbf{Heuristic 5:} The amount of asset sold in $T_{A2}$ must be within $90\% \sim 110\%$ of the amount bought in $T_{A1}$. If the sandwich attack is perfectly executed without interference from other market participants, the amount sold in $T_{A2}$ should be precisely equal to the amount purchased in $T_{A1}$. According to our empirical data~\SandwichExtractedPerfect(\SandwichExtractedPerfectPercentage) sandwich attacks we detect are ``perfect''. We further relax this constraint to cover $\pm 10\%$ slippage, thus finding~\SandwichExtractedImperfect(\SandwichExtractedImperfectPercentage) additional imperfect profitable sandwich attacks.
\end{itemize}

\subsubsection{Empirical Results}
In total, we identify~\SandwichExtractedNumEOA Ethereum user addresses and~\SandwichExtractedNumSC smart contracts performing~\SandwichExtractedNumAttacks sandwich attacks on~\PlatformsSandwich, with a total profit of~\SandwichExtractedTotalProfit (cf.\ Fig.~\ref{fig:sandwich-monthly-statistics}). Our heuristics do not find sandwich attacks on Curve, Swerve, and 1inch. Curve/Swerve are specialized in correlated, i.e., pegged-coins with minimal slippage. Despite the small market cap.\ ($<1\%$ of Bitcoin), SHIB is the most sandwich attack-prone ERC20 token with an adversarial profit of~$6.84$M USD.

We notice that \SandwichExtractedZeroGasPrice sandwich attacks (\SandwichExtractedPercentageZeroGasPrice) are privately relayed to miners (i.e., zero gas price), accumulating a profit of~\SandwichExtractedZeroGasPriceProfit. Sandwich attackers therefore actively leverage BEV relay systems (cf.\ Section~\ref{sec:bev-relayer}) to extract value.
We also observe that~\SandwichExtractedNotSameAddress of the attacks use different accounts to issue the front- and back-running transactions. 

\begin{figure}[tb]
    \centering

        
        \begin{subfigure}{\columnwidth}
        \includegraphics[width=\columnwidth]{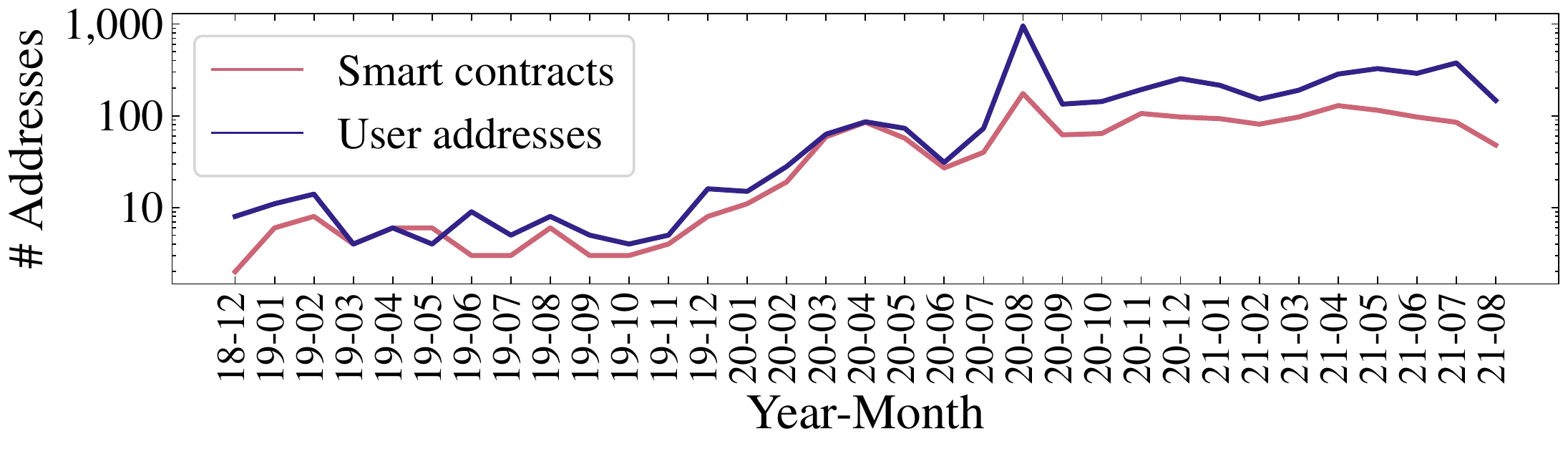}
        \caption{Number of active adversarial sandwich user addresses and smart contracts detected over time.}
        \label{fig:sandwich-attackers}
        \end{subfigure}
        
        \begin{subfigure}{\columnwidth}
        \includegraphics[width=\columnwidth]{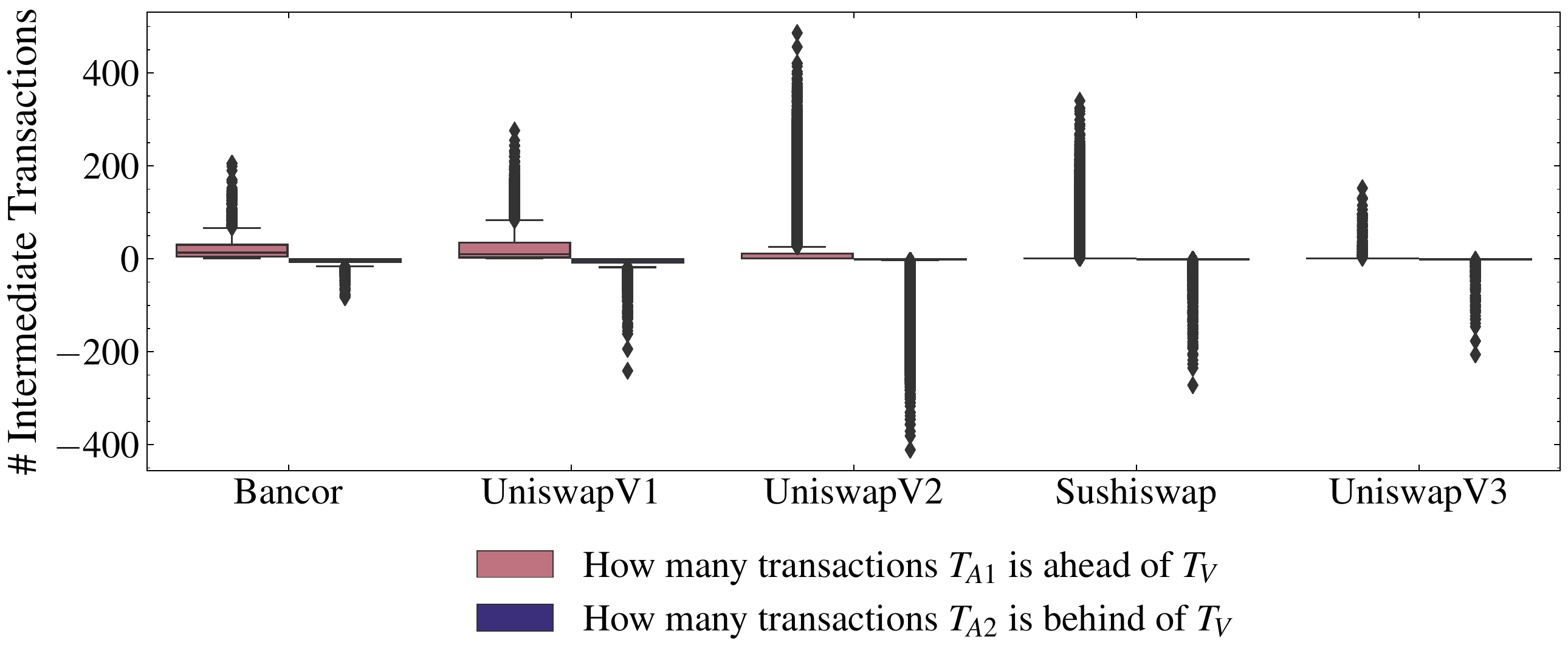}
        \caption{Relative position of sandwich transactions for profitable attacks.}
        \label{fig:sandwich-position}
        \end{subfigure}
        
        \caption{Extracted sandwich attacks, from block~\StartBlock (\StartDate) to block~\EndBlock (\EndDate).}
    \end{figure}

\point{Sandwich Transaction Positions}
A sandwich attack adversary typically attempts to position its transactions relatively close to the victim transaction. In practice, we observe multiple profitable sandwich attacks where the involved transactions are separated by more than $200$ intermediate transactions (cf.\ Fig.~\ref{fig:sandwich-position}), while no intermediate transaction (i.e., the front-running, victim, and back-running transactions are positioned one by one) is detected in~\SandwichPrivatelyRelayedPercentageCLose of the privately relayed sandwich attacks. We present the sandwich attack gas price distribution and adversarial strategies in Appendix~\ref{app:sandwich-attacks}.

\point{Extractable Profit} Zhou \etal~\cite{zhou2021high} estimate that under the optimal setting, the adversary can attack $\numprint{7793}$ Uniswap V1 transactions, and realize~$98.15$~ETH of revenue from block~8M to~9M. Based on our data, we estimate that only \SandwichExtractedPercentage (\SandwichExtractedHFTPaper) of the available extractable value was extracted.

\subsection{Fixed Spread Liquidations}\label{sec:fixed-spread-liquidation}
We observe two widely adopted liquidation mechanisms in the current DeFi ecosystem~\cite{qin2021empirical}. First, the fixed spread liquidation, used by Aave, Compound, and dYdX, allows a liquidator to purchase collateral at a fixed discount when repaying debt. Second, the auction liquidation, allows a liquidator to start an auction that lasts for a pre-configured period (e.g., $6$ hours~\cite{makerdao}). Competing liquidators bid on the (lowest possible) collateral price. In this section, we focus on the fixed spread liquidation, which allows to extract value in a single, atomic transaction. To perform a fixed spread liquidation, a liquidator $\mathcal{A}$ can adopt the following two strategies.

\begin{itemize}[left=0pt]
    \item \textbf{Block State Liquidation:} $\mathcal{A}$ detects a liquidation opportunity at block $B_i$ (i.e., after the execution of $B_i$). $\mathcal{A}$ then issues a liquidation transaction $T_A$, which is expected to be mined in the next block $B_{i+1}$. $\mathcal{A}$ attempts to \textbf{destructively front-run} competing liquidators with $T_A$.
    
    \item \textbf{Network State Liquidation:} $\mathcal{A}$ observes a transaction $T_V$, which will create a liquidation opportunity (e.g., an oracle price update that renders a collateralized debt liquidatable). $\mathcal{A}$ then \textbf{back-runs} $T_V$ with a liquidation transaction $T_A$.
\end{itemize}

\subsubsection*{Empirical Results} 
\begin{figure}[tb!]
    \centering
    \begin{subfigure}{\columnwidth}
    \includegraphics[width=\columnwidth]{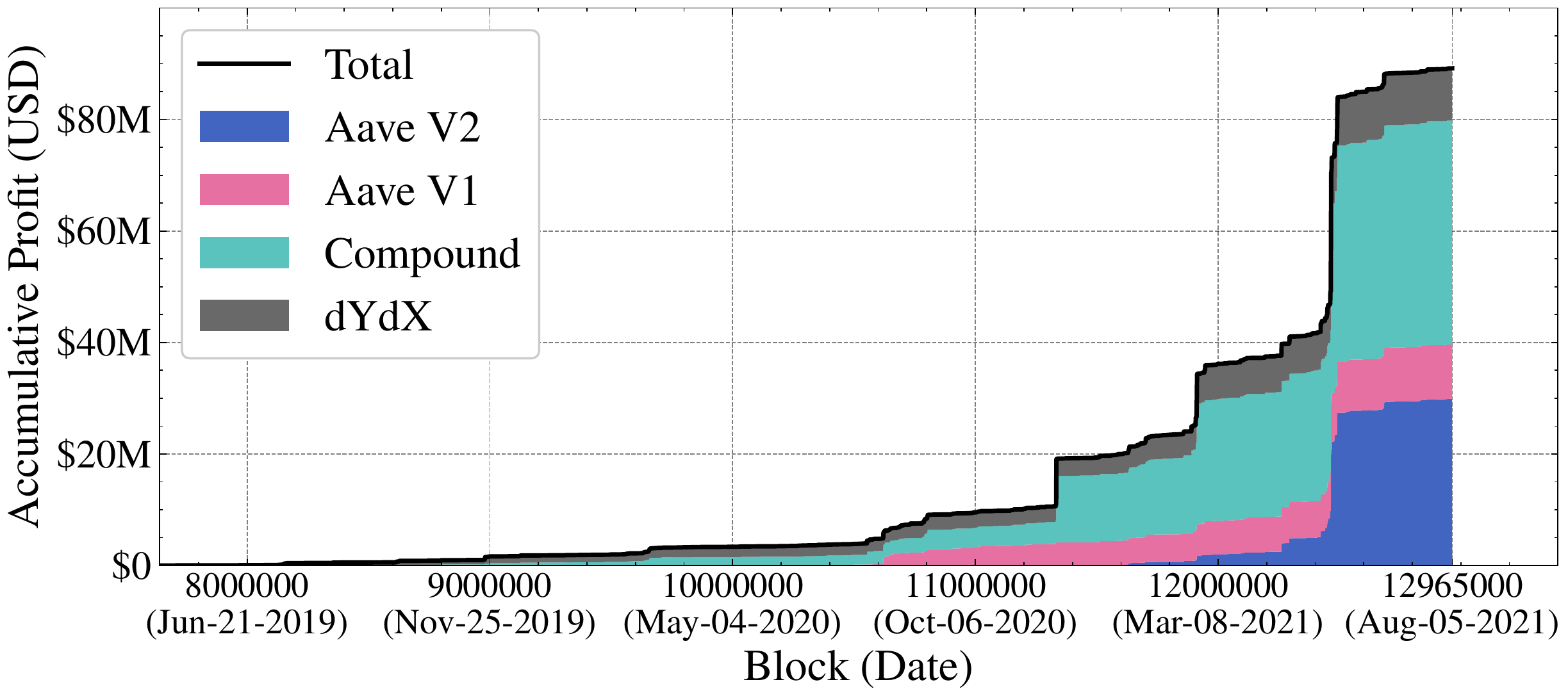}
    \caption{Accumulative profit of fixed spread liquidations.}
    \label{fig:accumulative-fixed_spread-liquidation-profit}
    \end{subfigure}
    \begin{subfigure}{\columnwidth}
    \includegraphics[width=\columnwidth]{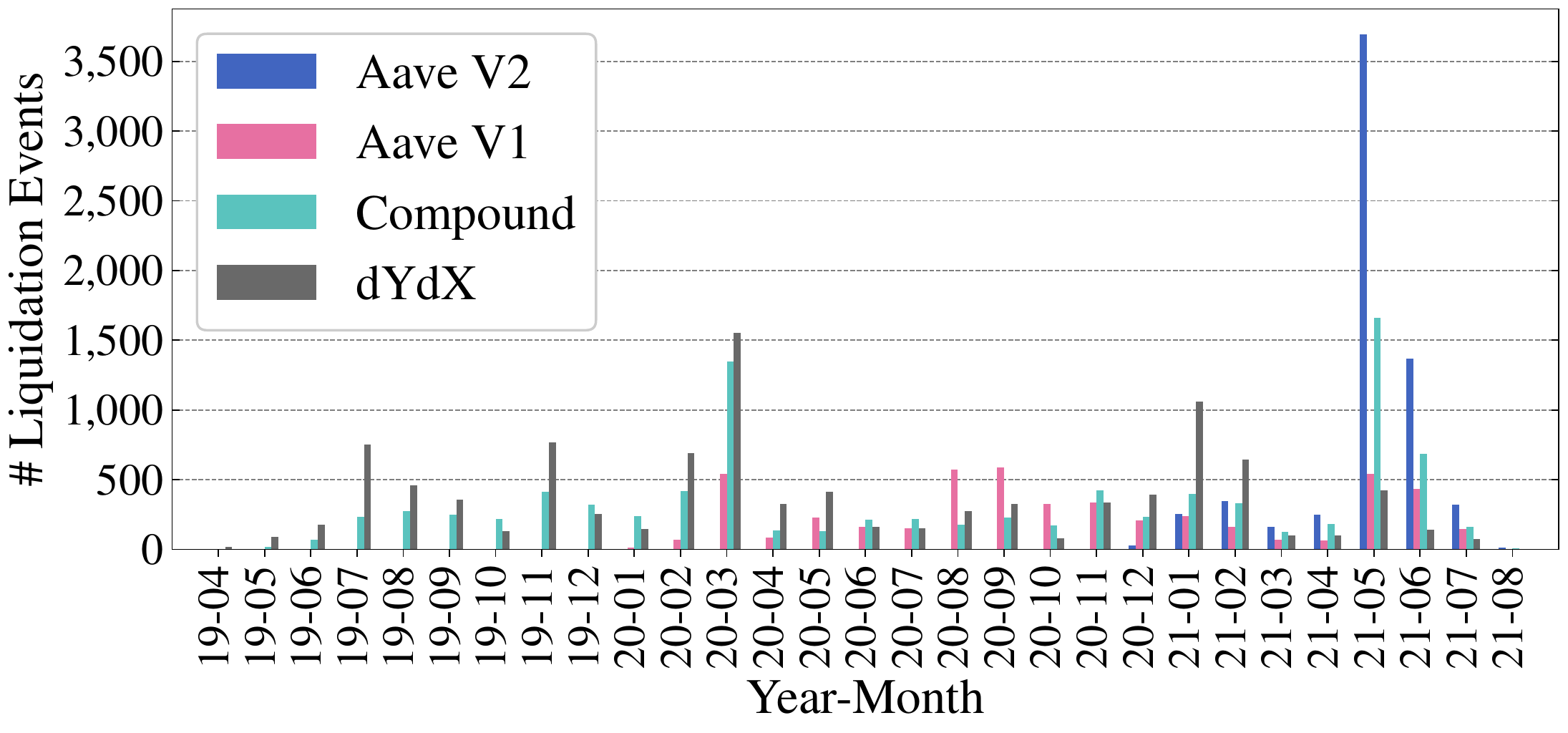}
    \caption{The monthly number of fixed spread liquidation events.}
    \label{fig:number-fixed_spread-liquidation}
    \end{subfigure}
    \caption{The number of liquidations increase in months where the ETH price collapses, e.g., in March,~2020 and May,~2021.}
\end{figure}

We collect all liquidation events on Aave (Version~$1$ and~$2$), Compound, and dYdX from their inception until block~\EndBlock (\EndDate). We observe a total of~\TotalNumFixedSpreadLiquidation liquidations, yielding a collective profit of~\TotalFixedSpreadLiquidationProfit over~\LiquidationCrawlingPeriod (cf.\ Fig.~\ref{fig:accumulative-fixed_spread-liquidation-profit} and~\ref{fig:number-fixed_spread-liquidation}). Note that we use the prices provided by the price oracles of the liquidation platforms to convert the profits to USD at the moment of the liquidation.

\begin{table}[b]
    \centering
    \caption{Extracting strategies of liquidators. Liquidators either back-run the price oracle updates, or front-run competing liquidation attempts. Most liquidations perform front-running.}
    \resizebox{0.8\columnwidth}{!}{%
    \begin{tabular}{@{}cccc@{}}
    \toprule
    \textbf{\begin{tabular}[c]{@{}c@{}}Liquidation Platform\end{tabular}} & \textbf{Front-running} & \textbf{Back-running} & \textbf{Total} \\ \midrule
    Aave V2 & \AaveVTwoFrontRunning & \AaveVTwoBackRunning & \AaveVTwoLiquidations \\
    Aave V1 & \AaveVOneFrontRunning & \AaveVOneBackRunning & \AaveVOneLiquidations \\
    Compound & \CompoundFrontRunning & \CompoundBackRunning & \CompoundLiquidations \\
    dYdX & \dYdXFrontRunning & \dYdXBackRunning & \dYdXLiquidations \\ \midrule
    \textbf{Total}  & \TotalLiquidationFrontRunning & \TotalLiquidationBackRunning & \TotalNumFixedSpreadLiquidation
    \\ \bottomrule
    \end{tabular}%
    }
    \label{tab:liquidation-strategies}
\end{table}

\point{Ordering Strategies}
To distinguish between a front- or back-running liquidation, we observe that a front-running liquidation at block $B_i$ necessarily requires a borrowing position to be liquidatable at block $B_{i-1}$. If the borrowing position is not liquidatable at block $B_{i-1}$, the liquidator is acting after a price oracle update in block $i$, which corresponds to a back-running liquidation. Therefore, for each of the~\TotalNumFixedSpreadLiquidation liquidations that we observe on block $B_i$, we test whether the borrowing position was liquidatable at block $B_{i-1}$. If this test resolves to true, we classify the liquidation as front-, otherwise as back-running (cf.\ Table~\ref{tab:liquidation-strategies}). Given~\TotalNumFixedSpreadLiquidation liquidations, we find that front-running is the dominating strategy accounting for~\LiquidationFrontRunningPercentage of all liquidations. Among the~\TotalNumFixedSpreadLiquidation liquidations, we identify~\TotalLiquidators unique liquidators by address. We find that~\FrontRunningLiquidators liquidators follow the front-,~\BackRunningLiquidators back-running and the remaining~\MixedLiquidators liquidators adopt a mixed strategy.

\begin{figure}
    \centering
    \includegraphics[width=\columnwidth]{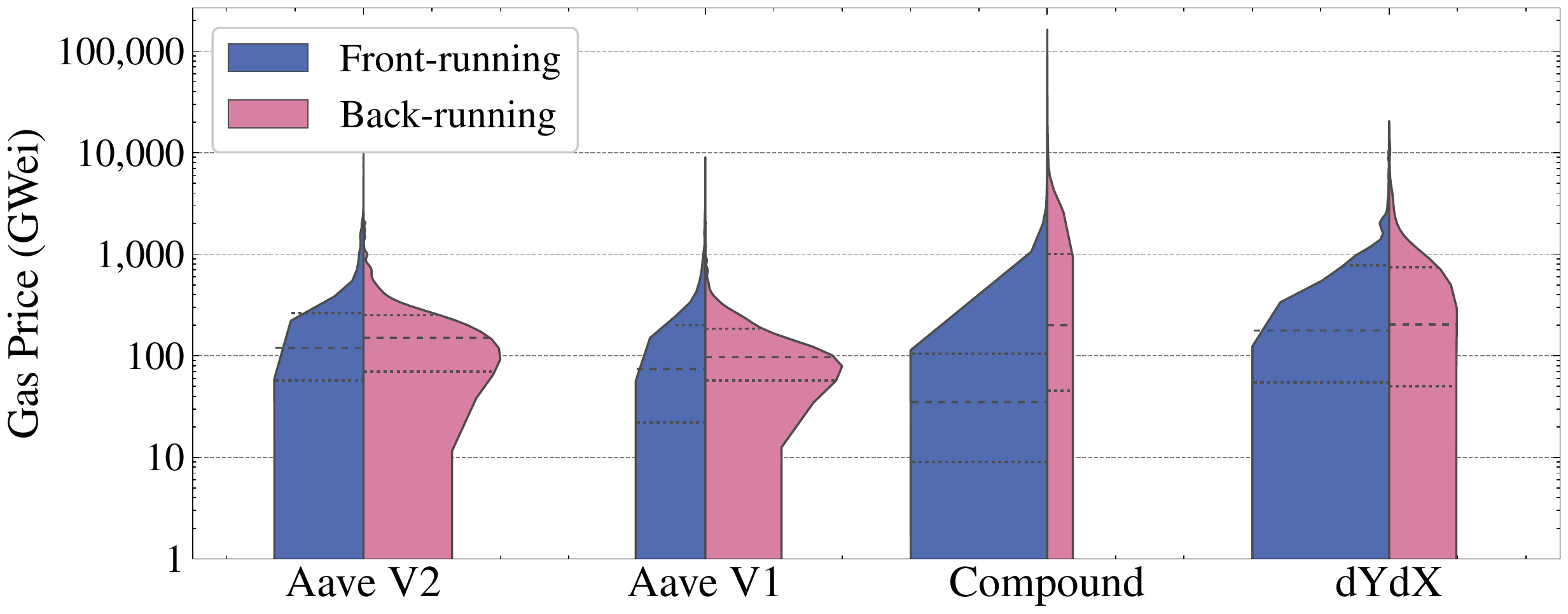}
    \caption{Transaction fee distributions of front- and back-running liquidations (transactions with zero gas price are excluded). The back-running liquidations pay a higher average gas price, due to the internal back-running concept.}
    \label{fig:liquidation-gasprice}
\end{figure}

\point{Liquidation Gas Prices} We identify~\ZeroGasPriceLiquidation transactions (\ZeroGasPriceLiquidationPercentage) with zero gas price out of the~\TotalNumFixedSpreadLiquidation liquidation events, implying that liquidators relay liquidation transactions to miners privately without using the P2P network. These privately relayed transactions yield a total profit of~\ZeroGasPriceLiquidationProfit. We visualize the gas price distributions in Fig.~\ref{fig:liquidation-gasprice}. Surprisingly, we notice that the back-running liquidations pay a higher gas prices on average. We find that this is because the liquidators tend to wrap the price oracle update action and liquidation into one (high-priority) transaction, which we term an \emph{internal back-running transaction}. The internal back-running transactions are typically set with a high gas price to prevent them from being front-run by competing liquidators.

\subsection{Arbitrage}\label{sec:arbitrage}

\begin{figure*}[tb]
    \centering
    \includegraphics[width=\textwidth]{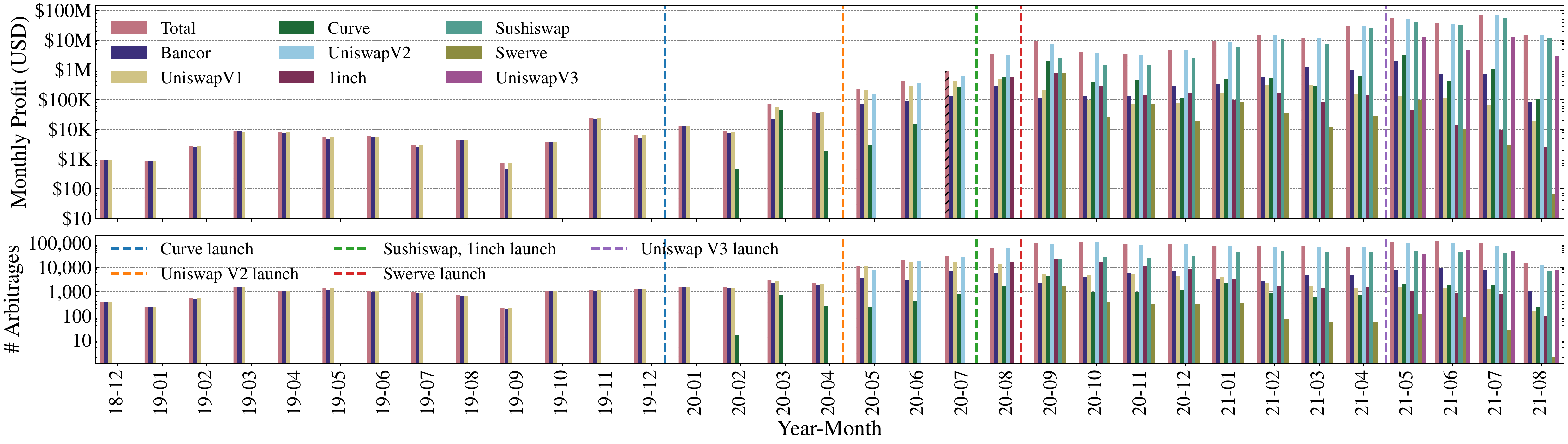}
    \caption{Monthly arbitrage statistics from block~\StartBlock (\StartDate) to block~\EndBlock (\EndDate).}
    \label{fig:arbitrage-monthly-statistics}
\end{figure*}




Arbitrage describes the process of simultaneously selling and buying assets in different markets in order to profit from the market price differences. Arbitrage helps to promote market efficiency and is typically considered benign. To perform an arbitrage, DeFi traders/miners monitor new blockchain state changes and execute an arbitrage if the expected revenue of synchronizing the prices on two markets exceeds the expected transaction costs. An arbitrage trader can choose among the following strategies to perform arbitrage:

\begin{itemize}[left=0pt]
    \item \textbf{Block State Arbitrage:} The arbitrage trader can choose to only monitor the confirmed blockchain states. Once a new block $B_i$ is received, the trader attempts to destructively front-run all other market participants at $B_{i+1}$.
    \item \textbf{Network State Arbitrage:} A trader can listen on the network layer to detect a ``large'' pending trade, which is likely to ``greatly'' change the asset price on one exchange. The trader then attempts to back-run this exchange transaction with an arbitrage transaction. 
\end{itemize}

\subsubsection{Heuristics}
We use $s$ to denote a swap action which sells $in(s)$ amount of the input asset $IN(s)$ to purchase $out(s)$ amount of the output asset $OUT(s)$. We apply the following heuristics to find extracted arbitrages on \Platforms.

\begin{itemize}[left=0pt]
\item \textbf{Heuristic 1:} All swap actions of an arbitrage must be included in a single transaction, implicitly assuming that the arbitrageur minimizes its risk through atomic arbitrage.
\item \textbf{Heuristic 2:} Arbitrage must have more than one swap action.
\item \textbf{Heuristic 3:} The $n$ swap actions $s_1, \ldots, s_n$ of an arbitrage must form a loop. The input asset of any swap action must be the output asset of the previous action, i.e., $IN(s_i) = OUT(s_{i-1})$. The first swap's input asset must be the same as the last swap action's output asset, i.e., $IN(s_0) = OUT(s_n)$.
\item \textbf{Heuristic 4:} The input amount of any swap action must be less than or equal to the output amount of the previous action, i.e., $in(s_i) \leq out(s_{i-1})$.
\end{itemize}


\subsubsection{Empirical Results}
From the~\StartDate to the~\EndDate, we identify~\ArbitrageExtractedNumEOA user addresses and~\ArbitrageExtractedNumSC smart contracts performing~\ArbitrageExtractedNumAttacks arbitrage trades on~\Platforms, amounting to a total profit of~\ArbitrageExtractedTotalProfit. We find that~\ZeroGasPriceArbitrage~arbitrage transactions (\ZeroGasPriceArbitragePercentage) are privately relayed to miners, representing~\ZeroGasPriceArbitrageProfit of extracted value. All detected arbitrage trades are executed using smart contracts.

\point{Arbitrage statistics}
To gain more insights on arbitrage, we classify the transactions according to the number of platforms and markets involved (cf.\ Table~\ref{tab:arbitrage_stats}). Most traders prefer simple strategies that only involve~$2$ or~$3$ markets (aka.\ two-point arbitrage and triangular arbitrage). Less than~$3\%$ of the transactions execute strategies with more than four markets. We, for example, find that one transaction combines two arbitrage into one to save gas costs\footnote{In the transaction \abbrEtherscanTx{0x07729d7826e2335a88ac1ae23aa9463a3183c6dc6e7a7ba485c244f473a9be87}, the trader executes the following arbitrage: WETH $\rightarrow$ BOXT $\rightarrow$ UNI $\rightarrow$ USDT $\rightarrow$ USDN $\rightarrow$ UNI $\rightarrow$ WETH. This strategy consists of two triangular arbitrages: \textit{(i)} WETH $\rightarrow$ BOXT $\rightarrow$ UNI $\rightarrow$ WETH; \textit{(ii)} UNI $\rightarrow$ USDT $\rightarrow$ USDN $\rightarrow$ UNI}. Such optimizations may yield a higher profit while riskier because the more markets involved, the more competitors must be front-run. ETH, USDC, USDT, and DAI are involved in~$99.91\%$ of the detected arbitrages.


\begin{table}[tb]
\centering
\caption{Statistics of the profitable arbitrage trades we detect. Over~$90\%$ synchronize the prices across~$2$ or~$3$ markets.}
\resizebox{\columnwidth}{!}{%
\begin{tabular}{c|rrrr|r}
\toprule
\diagbox{\textbf{\# markets}}{\textbf{\# platforms}} &            \multicolumn{1}{c}{$1$} &            \multicolumn{1}{c}{$2$} &   \multicolumn{1}{c}{$3$} &       \multicolumn{1}{c}{$\geq4$} &      \multicolumn{1}{|c}{\textbf{Total}} \\
\midrule
$2$ & $\numprint{8220}$ ($0.7\%$) & $\numprint{452148}$ ($39.3\%$) & N/A &  N/A & $\numprint{460368}$ ($40.0\%$) \\
$3$ & $\numprint{333039}$ ($28.9\%$) & $\numprint{235878}$ ($20.5\%$) & $\numprint{16431}$ ($1.4\%$) &  N/A & $\numprint{585348}$ ($50.8\%$) \\
$4$ & $\numprint{42816}$ ($3.7\%$) & $\numprint{28963}$ ($2.5\%$) &  $\numprint{7497}$ ($0.7\%$) & $\numprint{16}$ ($0\%$) & $\numprint{79292}$ ($6.9\%$) \\
$5$ & $\numprint{9460}$ ($0.8\%$) & $\numprint{6996}$ ($0.6\%$) & $\numprint{588}$ ($0.1\%$) &  $\numprint{70}$ ($0\%$) & $\numprint{17114}$ ($1.5\%$) \\
$\geq6$ & $\numprint{2693}$ ($0.2\%$) & $\numprint{5292}$ ($0.5\%$) & $\numprint{1308}$ ($0.1\%$) &  $\numprint{33}$ ($0\%$) & $\numprint{9326}$ ($0.8\%$) \\\midrule
\textbf{Total}  & $\numprint{396228}$ ($34.4\%$) & $\numprint{729277}$ ($63.3\%$) & $\numprint{25824}$ ($2.2\%$) & $\numprint{119}$ ($0\%$) & \ArbitrageExtractedNumAttacks ($100\%$) \\
\bottomrule
\end{tabular}
}
\label{tab:arbitrage_stats}
\end{table}

\begin{figure}[bt!]
    \centering
    \includegraphics[width=\columnwidth]{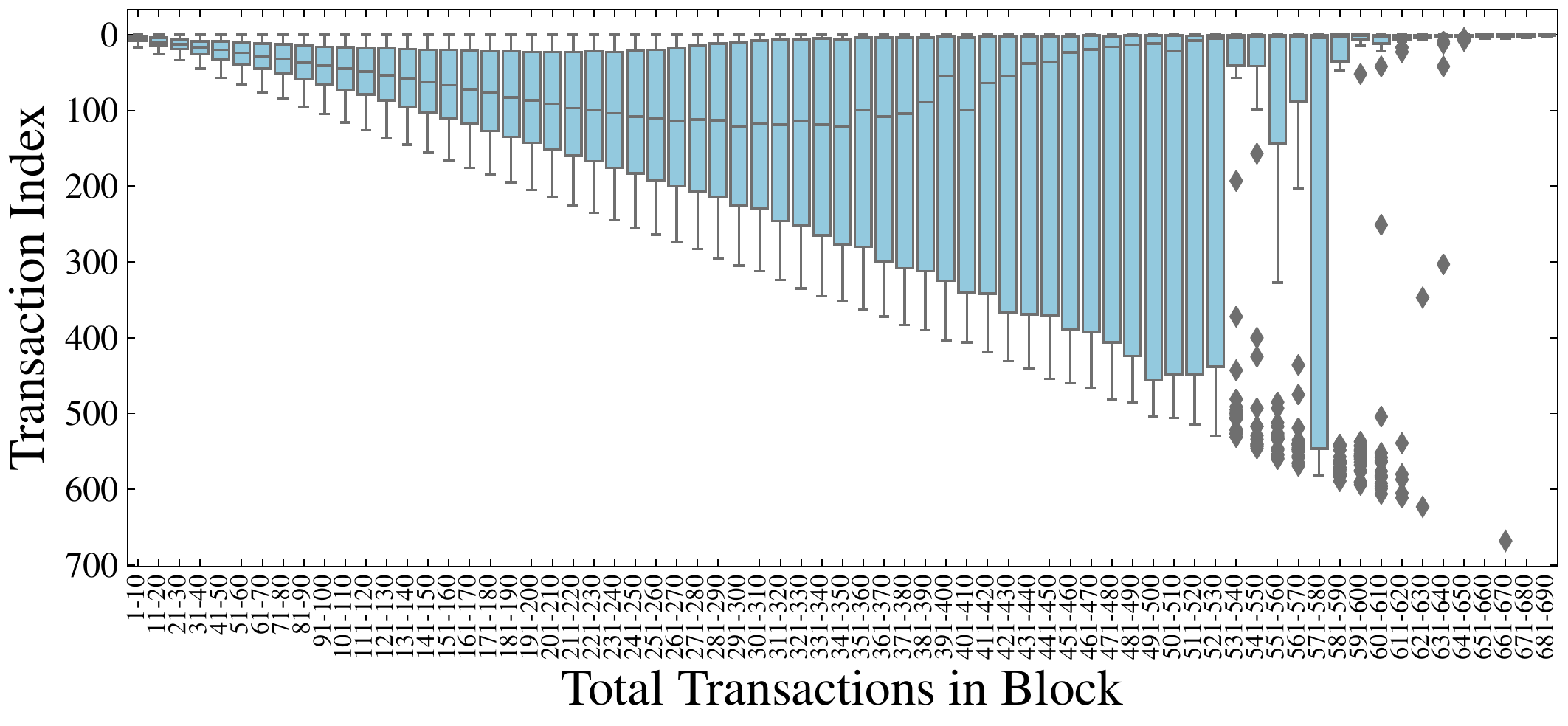}
    \caption{Transaction index distribution of all arbitrages we detect.}
    \label{fig:arbitrage_transaction_index}
\end{figure}

\point{Arbitrage transaction positions}
By visualizing the arbitrage transaction positions in blocks (cf.\ Fig.~\ref{fig:arbitrage_transaction_index}), we find that a large number of profitable trades are surprisingly positioned at the end of the blocks. We would have expected that the arbitrage transactions are competitive and perform destructive front-running with higher gas prices. For example, one of the most profitable arbitrage transactions\footnote{In the transaction \abbrEtherscanTx{0x2c79cdd1a16767e90d55a1598c833f77c609e972ea0fa7622b70a67646a681a5}, the trader first swaps~$400$ ETH for~$1040$ COMP on Uniswap v2, then swaps~$1040$ COMP for~$476$ ETH on Sushiswap, realizing a revenue of~$76$ ETH.} we detect is positioned at index~$141$ out of~$162$ transactions in this block. Our data hence supports the hypothesis that arbitrageurs are performing back-running on the network layer. To further confirm this hypothesis, we re-execute all arbitrage transactions at the top of blocks (i.e., upon the previous block state). If a transaction is a block state arbitrage, then the execution should remain profitable. We find that~\ArbitrageBackRunning of the arbitrage transactions are no longer profitable, which indicates that these transactions perform back-running because the arbitrage opportunity appears in the same block as the arbitrage transaction.

\subsection{Clogging}
We observe the practice of blockchain clogging by issuing simultaneously many transactions to intermediately increase the costs of writing to the blockchain. We identify various apparent purposes, such as attacking gambling protocols and mass token transfers (cf.\ Appendix~\ref{app:clogging} for quantitative details).

\subsection{Limitations}\label{sec:limitations}
We proceed to outline the main limitations of our measurements. Notably, as we focus on sandwich attacks, liquidations, and arbitrage, we do not capture all possible sources of BEV. We, however, believe that our methodology can be applied to other BEV sources. Then, for each BEV source, given that we apply custom heuristics, those heuristics have limitations themselves, which may result in false negatives. For instance, Heuristic~$1$ from the sandwich attacks assumes, that all transactions must be mined in the same block. There may exist successful sandwich attacks across multiple blocks, which we do not capture and which may result in false negatives. Also, it could be that by chance two transactions are executed right before and after a supposed victim transaction. Yet, this is not necessarily an attack. As such, heuristics may also introduce false positives into our findings. To reduce the potential inaccuracies of our heuristics, we attempt to tighten the heuristics to avoid overly reporting revenues. Summarizing, we do not have access to ground truth, which forces us to present our results as estimates only.

\section{Generalized Front-running: Transaction Replay}\label{sec:naivereplay}
\begin{figure}[tb]
    \centering
    \includegraphics[width=\columnwidth]{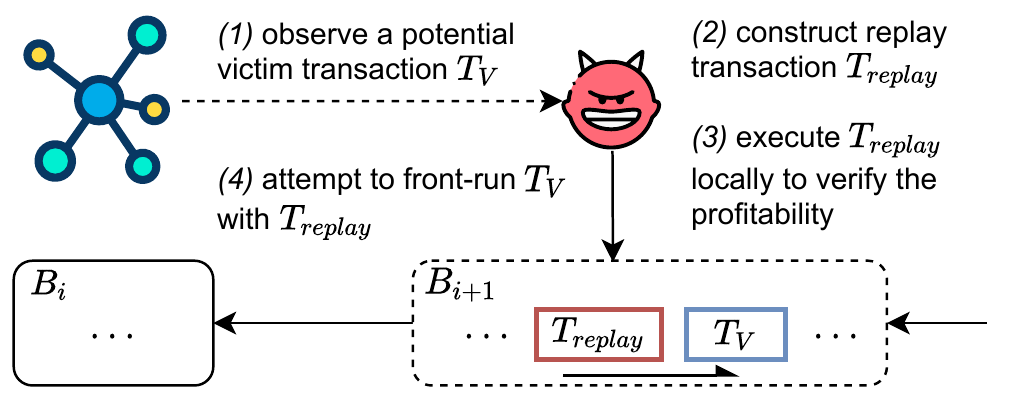}
    \caption{Overview of the transaction replay attack.}
    \label{fig:replay-attack-overview}
\end{figure}

We proceed to present an application-agnostic method for an adversary $\mathcal{A}$ to extract value by copying and replaying the execution logic of an unconfirmed victim transaction (cf. Fig.~\ref{fig:replay-attack-overview}). The high-level operations are as follows.

\begin{enumerate}
    \item $\mathcal{A}$ observes a victim transaction on the network layer;
    \item $\mathcal{A}$ constructs one or more replay transaction(s) to copy the execution logic of the victim transaction while diverting the revenue to an adversary-controlled account;
    \item $\mathcal{A}$ performs concrete validation of the constructed replay transaction(s) locally to emulate the execution result;
    \item if the local execution yields a profit, $\mathcal{A}$ attempts to \textbf{destructively front-run} the victim transaction.
\end{enumerate}

We classify a replay transaction $T_{replay}$ as profitable, if the native cryptocurrency (e.g., ETH) balance of $\mathcal{A}$ increases after the execution of $T_{replay}$, discounting the transaction fees. To measure profitability, we assume that $\mathcal{A}$ converts all the received assets (i.e., tokens) within an atomic transaction to the native cryptocurrency following the replay action.


\newcommand{\TransferRevenueToSender}{{\color{blue}\normalfont\texttt{TransferRevenueToSender}}\xspace}
\newcommand{\SpecifyBeneficiary}{{\color{blue}\normalfont\texttt{SpecifyBeneficiary}}\xspace}

\subsection{Algorithm}\label{sec:replay-algorithm}
Traders frequently implement profit-generating strategies (e.g., arbitrage) within smart contracts to perform complex operations atomically~\cite{qin2021attacking}. We however show that the following programming patterns expose a transaction to be replayable.
\begin{itemize}[left=0pt]
    \item \textbf{Sender Benefits:} The generated revenue is transferred to the transaction sender (cf.\ \TransferRevenueToSender in Listing~\ref{lst:naive_replay_examples}) without authentication.
    \item \textbf{Controllable Input:} The sender address is specified in the transaction input to receive the revenue (cf.\ \SpecifyBeneficiary in Listing~\ref{lst:naive_replay_examples}).
\end{itemize}

\subsubsection*{Replay Algorithm}
Generally, in a transaction $T$ on a smart-contract-enabled blockchain (cf.\ Eq.~\ref{eq:blockchain-transaction}), $sender$ represents the issuer of $T$, $value$ the amount of native cryptocurrency sent in $T$, and $input$ controls the contracts' execution\footnote{We ignore irrelevant fields (e.g., nonce).}. $sender$ is an authenticated field verified through the signature, and $input$ is arbitrarily amendable.
\begin{equation}\label{eq:blockchain-transaction}
    T = \{sender, value, input\}
\end{equation}

\begin{lstlisting}[float,floatplacement=H,label=lst:naive_replay_examples,language=Solidity, caption={Examples of the transaction replay algorithm patterns.}]
pragma solidity ^0.6.0;

contract Moneymaker {
  function TransferRevenueToSender() public {
    uint profit;
    // profiting logic omitted for brevity
    msg.sender.transfer(profit);
  }

  function SpecifyBeneficiary(address payable beneficiary) public {
    uint profit;
    // profiting logic omitted for brevity
    beneficiary.transfer(profit);
  }
}
\end{lstlisting}

We outline the replay logic in Algorithm~\ref{alg:naive_replay}. When observing a previously unknown transaction, the adversary constructs the replay transaction(s) by duplicating all the fields of the potential victim transaction but substitutes the original transaction sender address in the input data field with the adversarial address. 
An address in an Ethereum transaction input is encoded as a $20$-byte array\footnote{According to the Ethereum contract ABI specification~\cite{Contract43:online}, an address in the transaction data is left padded to $32$ bytes. However, the adversary is only concerned with the effective $20$ bytes when performing the substitution.}. Substitution is therefore efficient through a string replacement algorithm. The adversary then executes the replay transaction(s) locally upon the currently highest block. If the victim transaction conforms to the applicable patterns (i.e., sender benefits and controllable input), the execution of the replay transaction may yield a positive profit for the adversary, which can proceed with front-running the victim transaction. 

\begin{algorithm}[bt!]
\caption{Transaction Replay Algorithm.}
\small
\DontPrintSemicolon
\SetAlgoLined
\SetKwProg{Fn}{Function}{:}{end}
\SetKwFunction{FConstructReplay}{ConstructReplay}
\SetKwFunction{FSubstitute}{Substitute}
\SetKwProg{Alg}{Algorithm}{:}{end}
\SetKwFunction{ANaiveReplay}{TransactionReplay}
\KwIn{The current highest block $B_i$;\ the potential victim transaction $T_V$;\ the adversarial account address $\mathcal{A}$.
}\;


\Fn{\FConstructReplay{$T_V$, $\mathcal{A}$}}{
    $T.sender \leftarrow \mathcal{A}$\;
    $T.value \leftarrow T_V.value$\;
    $T.input \leftarrow$ substituting $T_V.sender$ in $T_V.input$ with $\mathcal{A}$\; 
    \KwRet{$T$}
}\;

\Alg{\ANaiveReplay{$T_V$, $\mathcal{A}$}}{
    $T_{replay} \leftarrow$\FConstructReplay{$T_V$, $\mathcal{A}$}\;
    Concretely Execute $T_{replay}$ upon block $B_i$\;
    \If{$T_{replay}$ is profitable}{Front-run $T_V$ with $T_{replay}$}
}
\label{alg:naive_replay}
\end{algorithm}

\subsection{Replay Evaluation}\label{sec:replay-evaluation}
We apply Algorithm~\ref{alg:naive_replay} to all the Ethereum transactions from block~\StartBlock (\StartDate)~to block~\EndBlock (\EndDate) capturing a total of~\TotalNaiveReplayTestTransactions~transactions over~\TimeDuration. We execute every constructed replay transaction at the position of the potential victim transaction and verify the profitability. Except for ETH, we consider all ERC20 tokens earned in the replay transactions as revenues. When a replay transaction yields a token revenue, we enforce an exchange transaction that converts the received token to ETH via on-chain Uniswap markets~\cite{httpsuni34:online}. We, therefore, measure the profitability entirely in ETH without the need for an external price oracle.
For simplicity of our analysis, we assume that the adversary pays $1$ Wei more than the victim transaction for the gas price of the replay and the potential exchange transaction (i.e., the minimal cost for a non-mining adversary to front-run). When measuring the profitability, we count the replay and exchange transaction fees as cost.

We perform our evaluation on a Ubuntu $20.04.1$ LTS machine with AMD Ryzen Threadripper $3990X$ ($64$-core, $2.9$ GHz), $256$ GB of RAM and $4\times2$ TB NVMe SSD in Raid $0$ conﬁguration. To execute a replay transaction in a past block, we download the blockchain state from an Ethereum full archive node running on the same machine. On average, generating a replay transaction and verifying its profitability takes~\ReplayPerformance (i.e., the time from observing a victim transaction to broadcasting the replay transaction). We remark that an adversary can achieve better performance by running the real-time replay attack inside an Ethereum client without downloading blockchain states from external sources.

\begin{figure}[tb]
\centering
\begin{subfigure}{\columnwidth}
\includegraphics[width=\columnwidth]{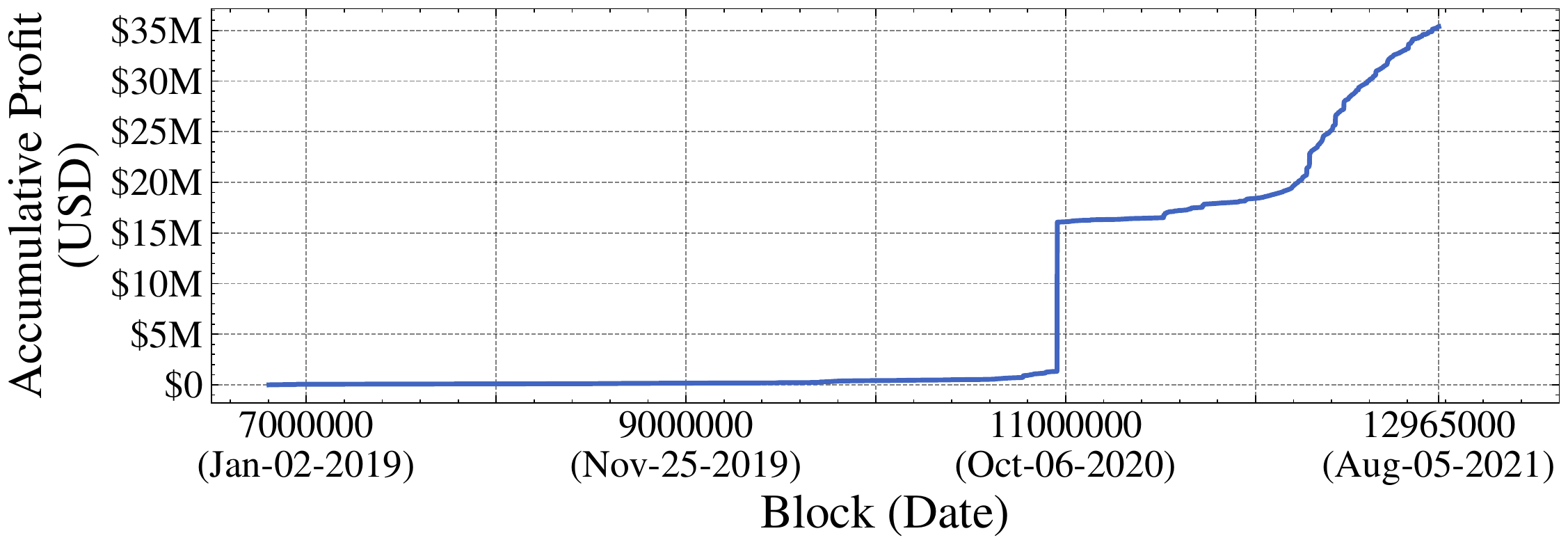}
\caption{Accumulative profit that can be extracted by replay attacks.}
\label{fig:transaction_replay_accumulative_profit}
\end{subfigure}
\begin{subfigure}{\columnwidth}
\includegraphics[width=\columnwidth]{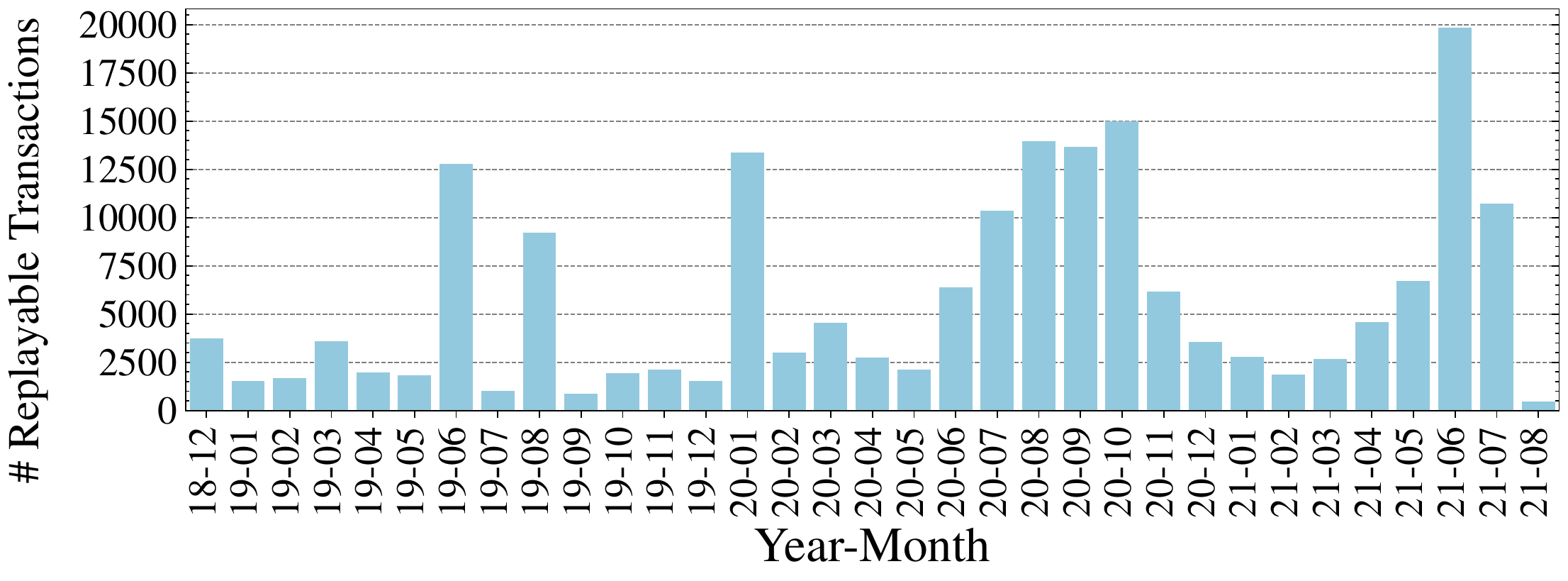}
\caption{Monthly number of replayable transactions.}
\label{fig:transaction_replay_number}
\end{subfigure}
\caption{Replay attacks amount to a profit of~\TotalReplayProfitUSD. We detect~\ReplayableTransctionInJuneTwentyTwenty replayable transactions in June,~2021 alone.}
\end{figure}

\subsubsection*{Results}
We find~\TotalReplayableTransactions profitable transactions (\TotalReplayableTransactionPercentage) that could have been replayed, accumulating to an estimated profit of~\TotalReplayProfit (\TotalReplayProfitUSD).
The most profitable replay transaction yields a profit of~\MostProfitableReplay. Apart from ETH, there are~\ReplayTransactionTokenNumber ERC20 tokens contributing a revenue of~\ReplayTransactionTokenRevenue in~\ReplayWithTokenTransactions transactions. Note that the ERC20 token revenue is higher than the total profit, because ETH is being used to purchase the ERC20 token in some transactions (recall that profit equals income minus expenses). Among all replayable transactions,~\BeneficiarysenderTransactions transactions follow the \emph{sender benefits} pattern, while the remaining~\SpecifyBeneficiaryTransactions transactions fall into the \emph{controllable input} category.

\begin{table}[b]
    \centering
    \caption{Required upfront ETH and average profit of replay.}
    \resizebox{\columnwidth}{!}{%
    \begin{tabular}{@{}ccc@{}}
    \toprule
    \multicolumn{1}{c}{\textbf{Required upfront capital $r$ (ETH)}} & \textbf{\# replay transactions} & \textbf{Average profit (ETH)} \\ \midrule
    $100 < r$ & $136$ & $2.48\pm8.05$ \\
    $10 < r \leq 100$ & $\numprint{2145}$ & $0.86\pm2.97$ \\
    $0 < r \leq 10$ & $\numprint{29372}$ & $0.21\pm3.93$ \\
    $r=0$ & $\numprint{156712}$ & $0.31\pm63.01$ \\ \bottomrule
    \end{tabular}%
    }
    \label{tab:upfrontcapitaldistribution}
\end{table}

We show the accumulative profit of the transaction replay attack in Fig.~\ref{fig:transaction_replay_accumulative_profit} along with the monthly number of replayable transactions in Fig.~\ref{fig:transaction_replay_number}. Notably, from block~\block{10954411} to~\block{10954419}, three transactions, which seem to exploit a smart contract vulnerability~\cite{DeFiDege78:online}, generate a total profit of over~\empirical{41,529~ETH}. We also observe a general uptrend in the number of replayable transactions since January,~2020.

In Table~\ref{tab:upfrontcapitaldistribution}, we show the distribution of the upfront ETH capital (i.e., the transaction value) required by the replay transactions, and outline the average profit. We find that~\empirical{$83.2\%$} of the replay transactions do not require upfront ETH, except the transaction fees. We notice that the replay profit is not directly correlated to the transaction value. \TotalTransactionsYieldsOverOneETH replay transactions yield a profit of more than one ETH, out of which~\TotalZeroTransactionsYieldsOverOneETH transactions are of zero-value.

We find~\ZeroGasPriceReplayableTransaction replayable transactions with zero gas price, representing a total value of~\ZeroGasPriceReplayableTransactionProfit. These privately relayed transactions hence are only replayable by mining adversaries or relay operators.
For the other transactions with positive gas price, in our evaluation, we assume that these transactions are at some point, prior to being mined, visible in the mempool. However, from the~22nd of December,~2020 to the~29th of December,~2020 (in prior to the emergence of BEV relay systems), we do not find~$13$ out of the~$1,156$ replayable transactions in our mempool (cf.\ Appendix~\ref{app:privatetransactions}). Our replay results may hence overestimate the replay potential by $1.12\%$.



\subsection{Real-Time Detection}\label{sec:real-time-detection}
Our previous replay results make use of historical on-chain data, and we extend this analysis with an investigation where we locally replay transactions in real-time from block~\RealtimeReplayStartBlock (\RealtimeReplayStartDate) to block~\EndBlock (\EndDate). To this end, we modify a go-ethereum client which connects to at most~$200$ peers. Following Algorithm~\ref{alg:naive_replay}, our client tests whether every received transaction from the P2P network is replayable. To avoid any doubt, our experiments remain local as we do not attempt to share our replay transactions.

\subsubsection*{Results} From a total of $\numprint{8206977}$ tested transactions, our real-time investigation find $166$ unique and non-conflicting transactions that are locally replayable. If we compare that number to the replayable candidates from on-chain data, within the same time-frame, we find $576$ unique (and non-conflicting) replayable transactions with a positive gas price. The discrepancy of those numbers indicates, that our node is insufficiently connected in the P2P network, and hence misses relevant replayable victims. We would welcome future work to use this metric as a success indicator of P2P network connectivity of an adversarial node.

The on-chain data moreover exposes $89$ replayable transactions with zero gas price. These transactions were likely mined through private agreements or a BEV relayer, and our real-time node naturally has no means to capture these transactions.

\subsection{Understanding Replayable Transactions}
The replay algorithm may act on any unconfirmed transaction without understanding its logic. To shed light on the nature of the replayable transactions, we cross-compare the~\TotalReplayableTransactions replayable transactions with the data from Section~\ref{sec:measurement}. We detect~\ReplayableLiquidationTransactions fixed spread liquidations (cf.\ Section~\ref{sec:fixed-spread-liquidation}) contributing a total profit of~\ReplayableLiquidationProfit, and~\ReplayableArbitrageTransactions arbitrages (cf.\ Section~\ref{sec:arbitrage}) contributing a total profit of~\ReplayableArbitrageProfit. These results suggest that the replay transactions capture a different set of profit-generating transactions than liquidations and arbitrage. In Appendix~\ref{app:replaycasestudy}, we provide a case study of replayable transactions. We find that two DeFi attacks are replayable, the Eminence exploit~\cite{DeFiDege78:online} and the bZx attack~\cite{qin2021attacking}.

\subsection{Naive Replay Protection}\label{sec:naive-replay-protection}
\newcommand{\Authentication}{{\color{blue}\normalfont\texttt{Authentication}}\xspace}
\newcommand{\MoveBeneficiary}{{\color{blue}\normalfont\texttt{MoveBeneficiary}}\xspace}



We proceed to present two simple methods that protect profitable transactions from being replayed by Algorithm~\ref{alg:naive_replay}.

    \point{(Insecure) Authentication} Authentication schemes are widely adopted in on-chain asset custody, e.g., when depositing assets into a smart contract wallet that can only be redeemed by an owner. Such schemes can also help to prevent simple replay attacks (cf.\ \Authentication in Listing~\ref{lst:naive_replay_protections}, Appendix~\ref{app:replay-protection}). When the authentication-enabled contract is invoked with an unauthorized address, the replay transaction execution is reverted. Such authentication method, however, does not remain secure against a more sophisticated replay algorithm.
    \point{Beneficiary Provision} To avoid a replay, the beneficiary address should not be specified in the transaction input and can instead be stored, for example, in the contract storage (cf.\ \MoveBeneficiary in Listing~\ref{lst:naive_replay_protections}, Appendix~\ref{app:replay-protection}).

The aforementioned methods mitigate the simple replay attacks. However, an adversary could go further in locally emulating a victim transaction, extract all emitted events and attempt to reconstruct its application layer logic. Specifically, the adversary can verify (e.g., given the heuristics of Section~\ref{sec:measurement}), if a transaction is an arbitrage or liquidation. A profitable transaction can then be constructed following the extracted application logic and parameters. We however remark that this replay method requires prior understanding of the specific application and therefore does not generalize further.

\subsection{Advanced Replay Protection}\label{sec:advanced-replay-protection}
A more robust replay protection mechanism requires that \emph{(i)} no entity besides the issuer can inspect the transaction and, \emph{(ii)} the miner can validate, but not view, the transaction.

Ironically, under strong trust assumptions, a BEV relayer, which we further discuss in the next section, may help to protect against replay attacks. The relayer, however, needs to be trusted and the miner must not perform replay attacks.

Fair ordering techniques~\cite{FairSequ5:online} (as further outline in Section~\ref{sec:bev-mitigation}) may also help to grant the original transaction issuer priority access to the blockchain. Unfortunately, state-of-the-art fair ordering techniques for permissionless blockchains are still vulnerable to well connected network layer adversaries.

A more elaborate alternative replay protection mechanism could be constructed with trusted hardware modules such as Intel SGX~\cite{costan2016intel}. Let's assume that miners are operating SGX enclaves ordering transactions within mined blocks. Traders could perform remote attestation to verify that the ordering enclave is following transparently outlined rules of inclusion. The trader can then establish an end-to-end encrypted TLS connection towards the miner enclave, and provide its transactions privately. The trader would be required to establish direct E2E-encrypted channels to all major miners/pools and concurrently send its transaction as in to avoid a replay attack. Unfortunately, in part due to DoS concerns, it is unclear whether miners would be willing to broadly open up their transaction ordering mining nodes to the public internet.

Also note that the approaches above are not immune to blockchain fork and reorganization attacks (which unfortunately are incentivised through BEV revenue, cf.\ Section~\ref{sec:insights}), as a transaction becomes public once its block is broadcasted.

\section{BEV Relayer and Auctions}\label{sec:bev-auctions}
Miners by default choose transactions from the mempool in a descending transaction fee order (e.g., gas price). The emerging BEV relayer, however, provide an additional transaction ``salesroom'': an trader propagates transactions to miners through a centralized relay system and shares the transaction profit with miners directly instead of paying transaction fees. In the following, we formalize an abstract BEV auction game capturing the P2P and the centralized BEV relayer model. We then quantitatively analyze how the introduction of BEV relayer impacts the P2P network and the consensus layer.

\subsection{BEV Relayer}\label{sec:bev-relayer}
\begin{figure}[bt!]
    \centering
    \includegraphics[width=\columnwidth]{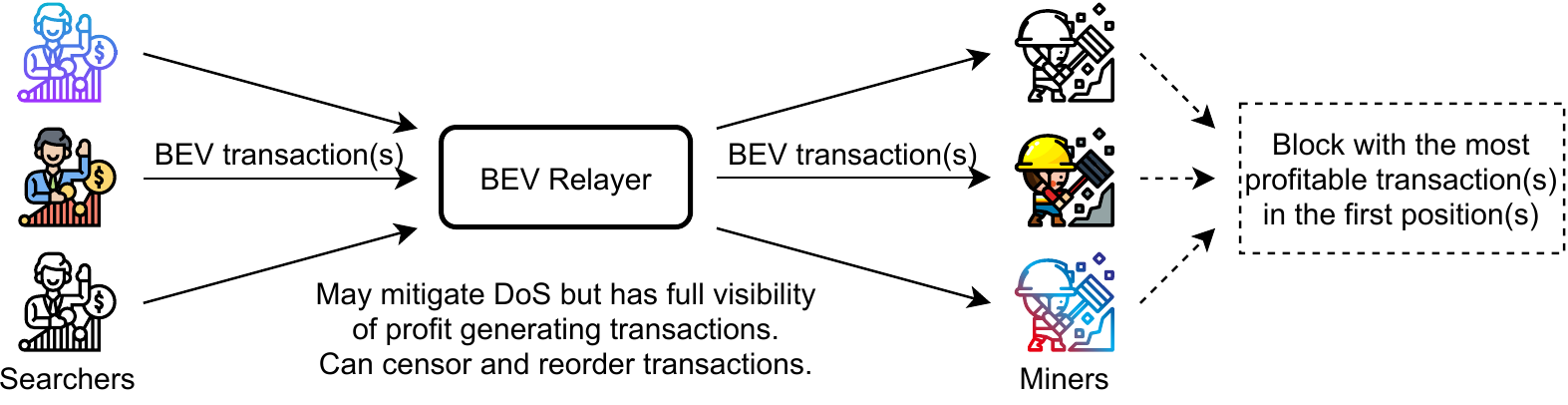}
    \caption{Architecture of a BEV relay mechanism, where a centralized and trusted server mediates between traders discovering BEV-extracting opportunities and miners.}
    \label{fig:bev-relayer}
\end{figure}

BEV relayers are centralized entities that provide a mediation service between traders seeking to extract BEV (so-called ``searchers'') and miners (cf.\ Fig.~\ref{fig:bev-relayer}). The relayer is a server, to which searchers submit one or multiple transactions (a bundle) that are then forwarded to the miners peered with the relayer. We observe that searchers perform sandwich attacks (cf.\ Section~\ref{sec:sandwich}) by packing the victim transaction and attack transactions into one bundle. The bundle fee mechanism guarantees that no transaction fee is paid if the transactions would fail. Miners operate an augmented client, which filters and positions the most profitable bundle(s) at the top of the next mined block. The BEV relay service is advertised to provide the following benefits: \textit{(i)}~The relayer claims not to publish BEV transactions. \textit{(ii)}~Searchers do not pay for failed transactions. \textit{(iii)}~Miners receive a share from the bundle revenue. \textit{(iv)}~P2P network congestion is claimed to be reduced. \textit{(v)}~Blockchain transaction fees are claimed to be reduced.


\subsection{BEV Auction Modeling}
We assume that a set of $n$ players $\{\mathcal{P}_0, \mathcal{P}_1, ..., \mathcal{P}_{n-1}\}$ compete for a BEV opportunity $\mathcal{O}$, which can be extracted through front- or back-running (cf.\ Section~\ref{sec:transaction-order-taxonomy}). We assume that if extracted, $\mathcal{O}$ yields a revenue of $\mathcal{R}_i(\mathcal{O})$ for player $\mathcal{P}_i$. Players may extract different values from the same opportunity depending on the extraction execution (e.g., the arbitrage paths and potentially sub-optimal parameters).

We call miners adopting the BEV relayer system ``relay miners'' and assume that relay miners control a hash-rate $\alpha$ of the total mining power. The remaining miners are denoted as ``P2P miners''. In this section, we assume that the BEV relayer honestly relays the transactions from players (i.e., searchers in Section~\ref{sec:bev-relayer}) without censoring or reordering transactions. We further assume that the relayer neither joins the BEV auction nor reveals any transaction to other players. The relay miners only pick the most profitable transactions(s) from the relayer system. We assume that the remaining block space is filled with the transactions from the P2P network sorted by the paid transaction fee. The P2P miners pick transactions solely from the P2P network in transaction fee descending order.

Every player $\mathcal{P}_i$ can participate in two optional auctions to extract $\mathcal{O}$. In the P2P auction, $\mathcal{P}_i$ broadcasts transactions in the P2P network. $\mathcal{P}_i$ places a publicly readable bid in the form of transaction fees. In the second auction, the relay auction, $\mathcal{P}_i$ does not broadcast transactions. Instead, $\mathcal{P}_i$ forwards crafted transactions to a centralized BEV relayer which forwards the transactions to relay miners. The relay miner is promised a share of the revenue of $\mathcal{R}_i(\mathcal{O})$, freely configurable by the player. We assume that players are rational, i.e., $\mathcal{P}_i$ participates in an auction \emph{iff} the expected payoff is positive. In the following, we use the term player and bidder interchangeably.

\point{P2P Auction (PA)}
The P2P auction is a first-price all-pay auction~\cite{baye1996all}, where the bidder only realizes a profit when its transaction is executed in the intended future block position. If the bidder's transaction does not execute at the intended position, upon block inclusion the bidder remains liable to a pay a transaction fee, or may realize a sub-optimal revenue.

We assume that $\mathcal{P}_i$ adopts the strategy $\mathcal{S}_i$ in the P2P auction, which provides a winning probability of $\operatorname{Pr}^\text{PA}(\mathcal{O}, \mathcal{S}_i)$. We further assume that $\mathcal{S}_i$ and  $\operatorname{Pr}^\text{PA}(\mathcal{O}, \mathcal{S}_i)$ are prior knowledge of $\mathcal{P}_i$ obtained from past experience. We formalize the expected payoff of a P2P auction participation in Eq.~\ref{eq:p2p-payoff}.
\begin{equation}\label{eq:p2p-payoff}
    \mathbb{E}\left[u_i^{\text{PA}}\mid\alpha\right]=(1-\alpha)\operatorname{Pr}^\text{PA}(\mathcal{O}, \mathcal{S}_i)\mathcal{R}_i(\mathcal{O})-b_i^{\text{PA}}(\mathcal{O}, \mathcal{S}_i)
\end{equation}
$b_i^{\text{PA}}(\mathcal{O}, \mathcal{S}_i)$ is the transaction fee $\mathcal{P}_i$ is willing to pay to the miners. Note that we ignore that the transaction execution result may impact the transaction fee. In a front-running competition, $\mathcal{P}_i$ might issue multiple transactions to increase the transaction fee bid, $b_i^{\text{PA}}(\mathcal{O}, \mathcal{S}_i)$ denotes the last bid. Eq.~\ref{eq:p2p-payoff} shows that the existence of BEV relayers (i.e., $\alpha$) decreases the players' expected payoff in the P2P auction. Players may hence refrain from broadcasting the BEV transactions. We further analyze the network layer impact of BEV relayers in Section~\ref{sec:networkimpact}.

\point{Relay auction (RA)}
A BEV relay auction is a first-price sealed-bid auction~\cite{easley2012networks} as bidders do not pay transaction fees unless they win. Eq.~\ref{eq:payoff-relayer-auction} outlines the payoff for $\mathcal{P}_i$ in the relay auction, when a replay miner produces the next block.
\begin{equation}\label{eq:payoff-relayer-auction}
    u_i^{\text{RA}}=\left\{
    \begin{aligned}
    & \mathcal{R}_i(\mathcal{O})-b_i^{\text{RA}}(\mathcal{O}) &\text{if $\mathcal{P}_i$ wins the auction}\\
    & 0  &\text{otherwise}
    \end{aligned}
    \right.
\end{equation}
$b_i^{\text{RA}}(\mathcal{O})$ is the rebate bidders pay to the miner. Note that a rational bidder would only pay a fee inferior to the revenue that $\mathcal{O}$ yields, i.e., $b_i^{\text{RA}}(\mathcal{O}) < \mathcal{R}_i(\mathcal{O})$. 

\subsection{Incentive Compatibility of the Relay auction Participation}\label{sec:relayauction}
Under a rational setting, the relay auction payoff for $\mathcal{P}_i$ is non-negative (cf.\ Eq.~\ref{eq:payoff-relayer-auction}). This result implies that players are always encouraged to participate in the relay auction, regardless of the mining power of relay miners or other players' strategies. \cite{daian2020flash} proposes a discouragement hypothesis that in the P2P front-running competition, players are discouraged by the market leaders and hence exit the game. We claim that this discouragement hypothesis never stands in the relay auction due to the risk-free nature of the relay auction (under the honest relayer assumption). Therefore, given the same BEV opportunity, the relay auction leads to a more intense competition than the P2P auction.

Increasing $b_i^{\text{RA}}(\mathcal{O})$ renders $\mathcal{P}_i$ more likely to win, but provides less payoff to the player. In a first-price auction, $\mathcal{P}_i$ does not have a dominant strategy (a strategy that maximizes the payoff) without knowing the other players' strategies~\cite{easley2012networks}, which makes it challenging to reason about how $\mathcal{P}_i$ should bid. We hence simplify and  assume that the reward $\mathcal{R}_i(\mathcal{O})$ is independently drawn from the same uniform distribution, i.e.,  $\mathcal{R}_i(\mathcal{O})\sim U(0, \mathcal{R}_{\max})$. We assume that $n$ and $U(0, \mathcal{R}_{\max})$ are prior knowledge of $\mathcal{P}_i$\footnote{In practice, $\mathcal{P}_i$ can approximate $U(0, \mathcal{R}_{
\max})$ or any other hypothetical distribution from all the P2P auction transactions, which are public, and $n$ from the success rate of the previous relay auctions.}. Under this simplifying assumption, the Bayesian Nash equilibrium strategy of $\mathcal{P}_i$ is shown in Eq.~\ref{eq:bidstrategy}, with the expected revenue of the miner provided in Eq.~\ref{eq:expectedrevenue}. Proofs for Eq.~\ref{eq:bidstrategy} and~\ref{eq:expectedrevenue} can be found in~\cite{easley2012networks}.
\begin{equation}\label{eq:bidstrategy}
    b_i^{\text{PA}}(\mathcal{O}, \mathcal{S}_i)=\frac{n-1}{n}\mathcal{R}_i(\mathcal{O})
\end{equation}
\begin{equation}\label{eq:expectedrevenue}
\mathbb{E}\left[\max_i b_i^{\text{PA}}(\mathcal{O}, \mathcal{S}_i)\right]=\frac{n-1}{n+1}\mathcal{R}_{\max}
\end{equation}

Eq.~\ref{eq:bidstrategy} implies that a player should bid more (i.e., pay higher fees) when the number of players increases. Therefore, the relay miners earn more revenue under a Bayesian Nash equilibrium when there are more relay auction bidders (cf.\ Eq.~\ref{eq:expectedrevenue}). We have shown that a first-price setting ensures a non-negative payoff, which incentivises participation. We can conclude that the first-price relay auction leans toward allocating the vast majority of BEV to relay miners, which we call \emph{revenue concentration}. This concentration then aggravates the incentivizes miners have to perform attacks on the consensus layer, which endangers the blockchain security (cf.\ Section~\ref{sec:insights}).

\subsection{Network Impact of the BEV Relayer}\label{sec:networkimpact}
BEV relayers advertise to reduce the P2P network layer congestion from competitive trading bots. In this section, we proceed to analyze when players actually refrain from broadcasting BEV transactions on the P2P network due to the availability of BEV relayer. We first define the concept of a \emph{protogenetic opportunity} (cf.\ Definition~\ref{def:protogenetic-opportunity}).

\begin{definition}{(Protogenetic Opportunity)}\label{def:protogenetic-opportunity} A BEV opportunity $\mathcal{O}$ is protogenetic for a player $\mathcal{P}_i$, if $\mathbb{E}\left[u_i^{\text{PA}}\mid 0\right] > 0$, i.e., the expected reward is positive when the mining power adopting BEV relayers is zero.
\end{definition}

Protogenetic opportunities represent the transactions that $\mathcal{P}_i$ would broadcast to the P2P network, when there is no BEV relayer. To quantify the impact of BEV relayers on the P2P network, we empirically measure how many protogenetic BEV transactions could have been prevented from propagating in the P2P network due to the introduction of BEV relayers. 


Given BEV relayers, $\mathcal{P}_i$ participates in the P2P auction only when $\mathbb{E}\left[u_i^{\text{PA}}\mid\alpha\right] > 0$. Following Eq.~\ref{eq:p2p-payoff} and Def.~\ref{def:protogenetic-opportunity}, we claim that the BEV relayers prevent $\mathcal{P}_i$ from broadcasting a transaction extracting $\mathcal{O}$ when satisfying Eq.~\ref{eq:condition}.
\begin{equation}\label{eq:condition}
    \frac{1}{\operatorname{Pr}^{\text{PA}}(\mathcal{O}, \mathcal{S}_i)} < \underbrace{\frac{\mathcal{R}_i(\mathcal{O})}{b_i^{\text{PA}}(\mathcal{O}, \mathcal{S}_i)}}_{\text{revenue-fee ratio}} < \frac{1}{(1-\alpha)\operatorname{Pr}^{\text{PA}}(\mathcal{O}, \mathcal{S}_i)}
\end{equation}

Intuitively, for an opportunity $\mathcal{O}$, if the revenue-fee ratio is too low, a rational player $\mathcal{P}_i$ will not broadcast the transaction, no matter whether a BEV relayer exists or not. If the revenue-fee ratio is high, $\mathcal{P}_i$ may still want to take a risk and participate in the P2P auction. Therefore, the BEV relay system only helps to discourage the propagation when the transaction revenue-fee ratio falls into the middle-range specified in Eq.~\ref{eq:condition}, given the mining power of relay miners (i.e., $\alpha$).



\subsubsection*{Results} 
\begin{figure}[tb]
    \centering
    \includegraphics[width=\columnwidth]{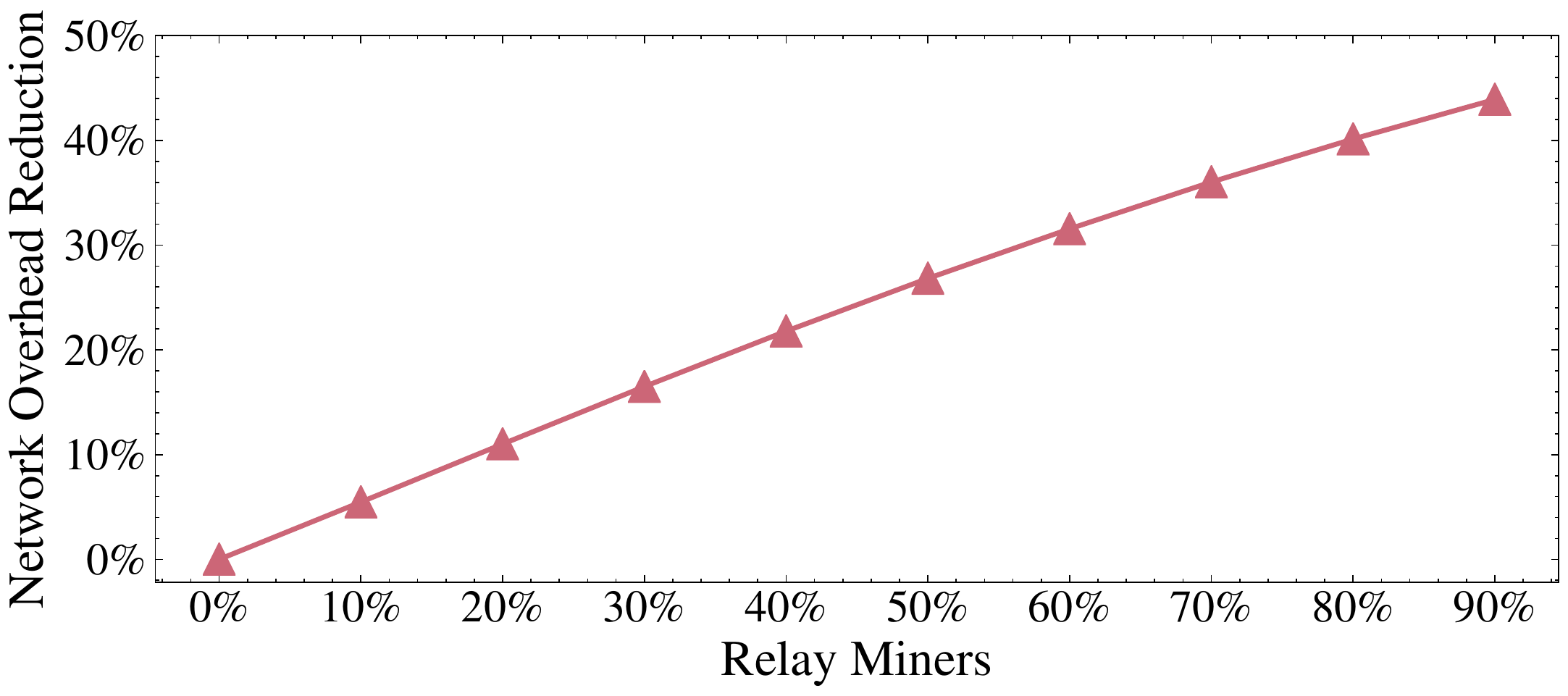}
    \caption{The percentage of the arbitrage transactions (cf.\ Section~\ref{sec:arbitrage}) that could have been prevented from broadcasting on the P2P network due to the introduction of BEV relayers.} 
    \label{fig:propagation-prevention-percentage}
\end{figure}

We measure the network impact of BEV relayers given the~\PositiveGasPriceArbitrage arbitrages with positive transaction fees in Section~\ref{sec:arbitrage}. Specifically, we calculate the revenue-fee ratio of every transaction and check if the ratio satisfies Eq.~\ref{eq:condition}. We are unaware of the winning probability of the players in the P2P auction (i.e., $\operatorname{Pr}^{\text{PA}}(\mathcal{O}, \mathcal{S}_i)$). Hence, for every transaction, we draw the value of $\operatorname{Pr}^{\text{PA}}(\mathcal{O}, \mathcal{S}_i)$ from a uniform distribution ranging from $10\%$ to $90\%$,  i.e., $\operatorname{Pr}^{\text{PA}}(\mathcal{O}, \mathcal{S}_i)\sim U(0.1, 0.9)$. We present the results under different relay mining power values in Fig.~\ref{fig:propagation-prevention-percentage}. Under $10\%$ relay miners, only~$5.5\%$ of the arbitrage transactions would be prevented from propagating in the P2P network. We show that even when~$90\%$ of the miners adopt BEV relayer systems, there are still~$56.1\%$ of the arbitrage transactions that would propagate in the P2P network.

\subsection{Privately Relayed Transactions}\label{sec:privatetransactions}
Transactions that are mined without appearing in the P2P network are referred to as privately relayed transactions. Besides BEV relayers, we notice that miners also reach agreements, e.g., with exchanges to mine privately propagated transactions. 
From the~22nd December,~2020 to 29th December,~2020 (prior to the emergence of BEV relayers), we identify~$\numprint{136143}$ privately relayed transactions out of a total of~$\numprint{8285218}$ ($1.64\%$). Detailed results are shown in Appendix~\ref{app:privatetransactions}.

\subsection{BEV Relayer Remarks}
Summarizing, our analysis provides the following novel and generic insights for smart contract enabled blockchains:
\begin{itemize}[left=0pt]
    \item BEV relayers aggravate consensus layer attacks by rendering MEV more competitive, yielding higher MEV opportunities and further incentivising miners to fork over MEV~\cite{zhou2021just}.
    \item Contrary to the suggestions of the practitioners community (e.g., \url{https://github.com/flashbots}), our results suggest that BEV relay mechanisms do not substantially reduce the P2P network overhead. That is despite the fact that a BEV relayer introduces an intermediary which increases the centralization of a permissionless blockchain.
\end{itemize}

\section{Security Insights of BEV}\label{sec:insights}

Previous studies~\cite{daian2020flash} have shown that the blockchain consensus is prone to time-bandit attacks, where miners deliberately fork and overwrite the main chain attempting to extract MEV (a subset of BEV). Zhou \etal~\cite{zhou2021just} point out that the time-bandit attacks are essentially equivalent to double-spending attacks, which can be captured by an MDP framework~\cite{gervais2016security}. When BEV is four times higher than the block reward, a financially rational miner with $10\%$ mining power is incentivized to fork the blockchain instead of performing honest mining.

\begin{figure}[bt]
    \centering
    \includegraphics[width=\columnwidth]{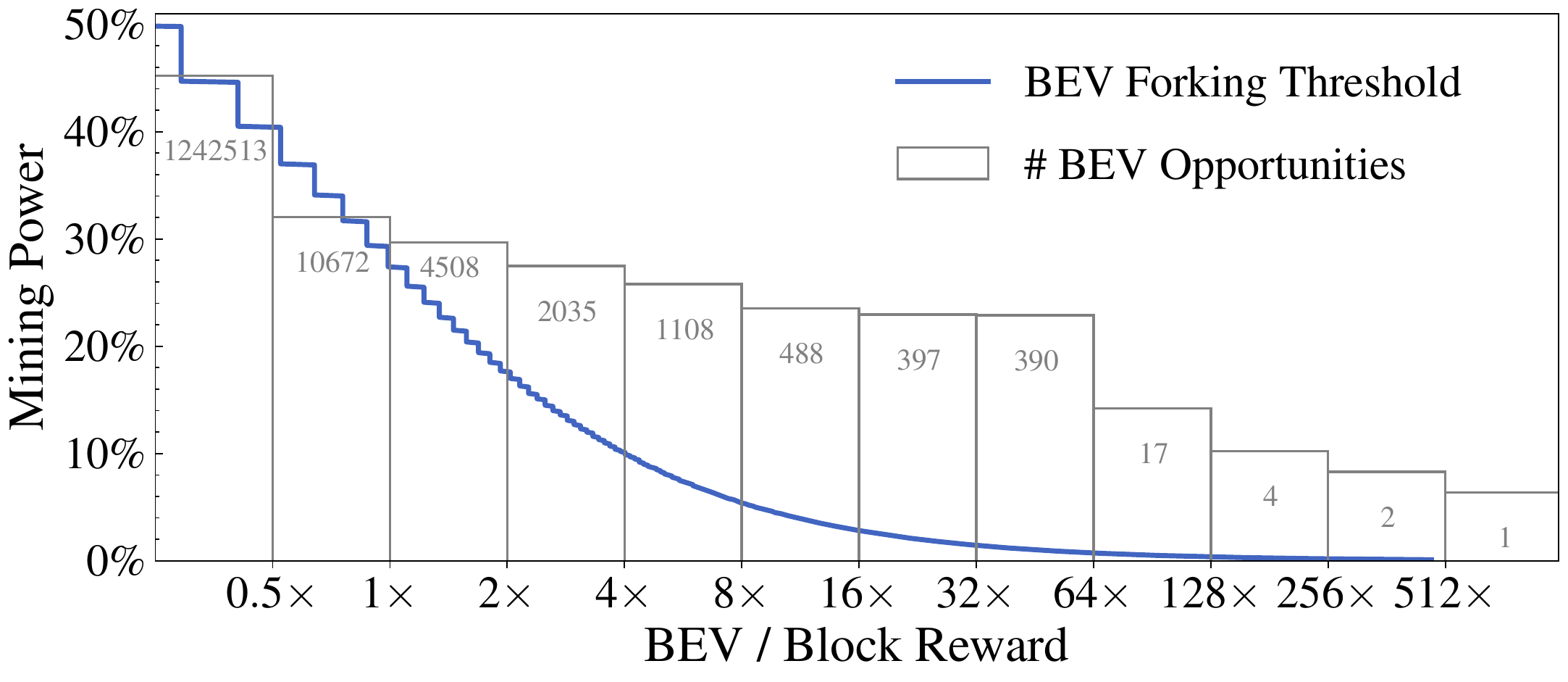}
    \caption{Minimum mining power on Ethereum that is incentivized to fork the chain to extract a BEV opportunity of $x\times$ the block reward (i.e., BEV forking threshold). We present the number of historical BEV opportunities per reward multiplier.}
    \label{fig:bev-security}
\end{figure}

Our measurements show that from the~\StartDate to the~\EndDate at least~\OverFourTimesBEVBlockNum blocks expose a BEV value of over four times the block reward plus transaction fees. The highest single-block BEV we find is \MaxSingleBlockBEVETH (\MaxSingleBlockBEVUSD) in block~\MaxSingleBlockBlockNum
~(\MaxSingleBlockBEVMultiple times the block reward plus transaction fees). This BEV opportunity could have incentivized a miner with only $0.1\%$ mining power to fork, which portrays the danger of drastic forking competition among BEV aware miners. To further understand empirically how the past BEV opportunities could have endangered the blockchain consensus security, we follow the MDP framework in~\cite{gervais2016security} and similar to Zhou \etal~\cite{zhou2021just} derive the \emph{BEV forking threshold} (cf.\ Fig.~\ref{fig:bev-security}). The BEV forking threshold captures the minimum mining power that is incentivized to fork the blockchain to extract a BEV opportunity of $x\times$ block reward. Fig.~\ref{fig:bev-security} further classifies each empirically identified BEV opportunity depending on its size with respect to the block reward.

BEV moreover provides miners an additional financial resources to perform bribery~\cite{bonneau2016buy} and undercutting attacks~\cite{carlsten2016instability}, where adversarial miners deliberately offer financial rewards (e.g., extractable BEV and transaction fee) on a forked chain to attract mining power. The revenue concentration objective of a BEV relayer further escalates the potential value that miners can extract, intensifying the risks of consensus layer forks.

BEV also causes congestion on the P2P network layer by attracting traders to heavily use the P2P network through many front- or back-running transactions. A congested P2P network, however, reduces communication throughput and latency, which was shown to increase the stale block rate, which in turn negatively affects consensus security~\cite{gervais2016security}.

\point{BEV relayer threats}
Throughout our analysis in Section~\ref{sec:bev-auctions}, we assume that BEV relayers and relay miners behave honestly. However, in reality, a relayer or miner may analyze and sell trader strategies in private. BEV relayer and miners can moreover replay profiting transactions (cf.\ Section~\ref{sec:naivereplay}). In a relay auction, all bids (i.e., the amounts of rebate that players pay to the miner) are visible to the relayer who can also manipulate the auction process. Knowing the highest bid, the relayer can for instance choose to bid a higher amount and win the auction. Through the use of multiple pseudonymous addresses, the relayer could deliberately pretend to lose auctions to deter manipulation detection. Such manipulation would lower the success rate of bidders and provide an illusion of a fierce competition, forcing bidders to raise their bids, aggravating the revenue concentration problem (cf.\ Section~\ref{sec:relayauction}). Finally, to the financial detriment of DeFi users, BEV relayers provide a risk-free approach to perform, for example, sandwich attacks.

\subsection{DeFi's Impact on BEV}\label{sec:defiimpactofbev}
DeFi is one of the most promising applications of permissionless blockchains. However, our empirical data from Section~\ref{sec:measurement}), intuitively suggests that the amount of extracted BEV grew with the overall DeFi TVL, hence clearly deteriorating blockchain security. Various DeFi attacks, including economic exploits~\cite{qin2021attacking} and sandwich attacks~\cite{zhou2021high}, are threatening DeFi users.

While BEV sources may appear benign from an application layer perspective (e.g., arbitrage synchronizes prices across different markets and liquidations help to secure debt), we claim that BEV should never be considered a desired ``feature'', and rather a design flaw. That is because BEV triggers transaction overhead and erodes the blockchain incentive mechanisms underpinned by the block reward and transaction fees.

\subsection{BEV Mitigation}\label{sec:bev-mitigation}
As long as the transaction executions remain transparent and the transaction order is unilaterally manipulable, the BEV challenge is likely to remain. Nevertheless, we observe several promising avenues towards reducing or mitigating BEV.

\point{Fair Ordering}
Kelkar \etal~\cite{kelkar2020order} formally define the concept of order-fairness and propose permissioned $\mathsf{Aequitas}$ protocols to order transactions fairly and were applied to DeFi~\cite{FairSequ5:online}. A variant of $\mathsf{Aequitas}$~\cite{kelkar2021order} extends order fairness for permissionless blockchains, yet a powerful network adversary retains an information asymmetry advantage to front-run slower victims.

\point{Application-Specific BEV Mitigation}
Previous works~\cite{zhou2021high} show that sandwich attacks can be mitigated if traders keep the trade sizes under the so-called minimum profitable victim input. Zhou \etal~\cite{zhou2021a2mm} propose the idea of exploiting a BEV opportunity atomically in the same transaction. For instance, when a trader performs an exchange on one market, an arbitrage opportunity might be created on another market. The trader can immediately execute an arbitrage following the exchange, which may yield an additional financial profit. Due to the atomicity of blockchain transactions, no adversary can extract the arbitrage profit. We can further imagine how BEV can be mitigated in lending protocols, if a price oracle update would atomically liquidate unhealthy debt position while paying out the liquidation revenue to a shared liquidity pool.

\section{Related Work}\label{sec:related-work}
Eskandir \etal~\cite{eskandari2019sok} are the first to introduce a front-running taxonomy for blockchains. While the authors focus on displacement, insertion and suppression front-running, we explicitly highlight the different side effects of adversarial front-running transactions, which therefore allows to differentiate between destructive or tolerating front-running. We moreover introduce the concept of back-running and show how these ordering strategies are used to extract value. We further identify the concept of an internal back-running transaction, where a transaction is atomically prepended with a ``high-priority'' transactions, such as a price oracle update.

Bonneau~\cite{bonneau2016buy} is the first to study bribery attacks in the context of Bitcoin-style consensus. With their seminal work, Daian \etal~\cite{daian2020flash} then introduce the concept of Miner Extractable Value, a specific financial source of bribing revenue. Through elaborate empirical data of the network layer, the authors show how competitive trading bots engage in front-running price gas auctions on the network layer. In this work, we offer quantifyable insights into the monetary value which traders have extracted through BEV, by analysing the historical blockchain data. We further capture regular and internal back-running, and propose the first practical transaction replay algorithm. Finally, we also model BEV relayer, which converts part of the public bidding game into a private relay auction.

Zhou \etal~\cite{zhou2021high} focus on the problem of sandwich attacks on AMM exchanges. The authors simulate, based on past blockchain data, how much revenue an adversary could have yielded theoretically from sandwich attacks. In this work, we measure the actual value extracted by sandwich adversaries, based on past blockchain data. Our data in Section~\ref{sec:sandwich} suggests, that only \SandwichExtractedPercentage (\SandwichExtractedHFTPaper) of the available extracted sandwich attack value was extracted. 

Related work captures extensively blockchain security through various models and quantification efforts. The most commonly captured attacks are selfish mining~\cite{eyal2014majority}, double-spending~\cite{gervais2016security}, bribery~\cite{bonneau2016buy}, and undercutting attacks~\cite{carlsten2016instability}. Zhou \etal~\cite{zhou2021just} quantify the value threshold at which MEV would incentivize miners to fork the blockchain based on optimal adversarial strategies given by an MDP. Based on this model, we empirically show the extent to which BEV could have endangered the blockchain consensus layer.

\section{Conclusion}\label{sec:conclusion}
In this paper we shed light on the practices of obscure and predatory traders on blockchains. We provide empirical data for the state-of-the-art BEV, by notably studying past sandwich attacks and arbitrage on seven decentralized exchanges as well as liquidations on three lending platforms. To the best of our knowledge, we are the first to provide a generalized real-time replay trading algorithm. We alarmingly observe that the emerging BEV relayer endanger the blockchains' security. We hope that our work provides insights into the current practices, and further helps to improve DeFi and blockchain security.


\bibliographystyle{IEEEtran}
\bibliography{references.bib}

\appendices
\section{Additional empirical data}

\subsection{Sandwich attack}\label{app:sandwich-attacks}
\subsubsection{Monthly Statistics}
Table~\ref{tab:sandwich_stat} shows the detailed monthly statistics of the sandwich attacks on Ethereum. We observe an increase in the number of attacks and the number of adversarial addresses (user/smart contract) from~$2020$. In April~$2021$, we find~$\numprint{94956}$ attacks, of which $96.5\%$ occur on Uniswap V2.

\begin{table*}[hbt!]
\centering
\caption{Monthly statistics of the sandwich attacks on Ethereum.}
\resizebox{\textwidth}{!}{%
\begin{tabular}{lllllllllllllllllllllllllllllllllll}
\toprule
{} &   Total &   18-12 &   19-01 &  19-02 &  19-03 &  19-04 &  19-05 &  19-06 &  19-07 &  19-08 &  19-09 &  19-10 &  19-11 &  19-12 &  20-01 &  20-02 &  20-03 &  20-04 &  20-05 &  20-06 &  20-07 &  20-08 &  20-09 &  20-10 &  20-11 &  20-12 &  21-01 &  21-02 &  21-03 &  21-04 &  21-05 &  21-06 &  21-07 &  21-08 \\
\midrule
Num. of smart contracts  &    1069 &       2 &       6 &      8 &      4 &      6 &      6 &      3 &      3 &      6 &      3 &      3 &      4 &      8 &     11 &     19 &     59 &     85 &     57 &     27 &     40 &    175 &     62 &     64 &    106 &     97 &     93 &     81 &     97 &    129 &    115 &     97 &     85 &     48 \\
                         &         &    0.2\% &    0.6\% &   0.7\% &   0.4\% &   0.6\% &   0.6\% &   0.3\% &   0.3\% &   0.6\% &   0.3\% &   0.3\% &   0.4\% &   0.7\% &   1.0\% &   1.8\% &   5.5\% &   8.0\% &   5.3\% &   2.5\% &   3.7\% &  16.4\% &   5.8\% &   6.0\% &   9.9\% &   9.1\% &   8.7\% &   7.6\% &   9.1\% &  12.1\% &  10.8\% &   9.1\% &   8.0\% &   4.5\% \\
Num. of user addresses   &    2419 &       8 &      11 &     14 &      4 &      6 &      4 &      9 &      5 &      8 &      5 &      4 &      5 &     16 &     15 &     28 &     63 &     86 &     73 &     31 &     73 &    955 &    134 &    143 &    193 &    254 &    215 &    152 &    190 &    285 &    327 &    290 &    377 &    147 \\
                         &         &    0.3\% &    0.5\% &   0.6\% &   0.2\% &   0.2\% &   0.2\% &   0.4\% &   0.2\% &   0.3\% &   0.2\% &   0.2\% &   0.2\% &   0.7\% &   0.6\% &   1.2\% &   2.6\% &   3.6\% &   3.0\% &   1.3\% &   3.0\% &  39.5\% &   5.5\% &   5.9\% &   8.0\% &  10.5\% &   8.9\% &   6.3\% &   7.9\% &  11.8\% &  13.5\% &  12.0\% &  15.6\% &   6.1\% \\
Num. of detected attacks &  750529 &      52 &     756 &    495 &    229 &    365 &    745 &    896 &    589 &    375 &    184 &    295 &    479 &    621 &   2117 &   1337 &    962 &   1052 &   3138 &   5991 &  12527 &  23393 &  34306 &  54980 &  41659 &  48748 &  35996 &  44513 &  71218 &  94956 &  85095 &  90152 &  80977 &  11331 \\
                         &         &    0.0\% &    0.1\% &   0.1\% &   0.0\% &   0.0\% &   0.1\% &   0.1\% &   0.1\% &   0.0\% &   0.0\% &   0.0\% &   0.1\% &   0.1\% &   0.3\% &   0.2\% &   0.1\% &   0.1\% &   0.4\% &   0.8\% &   1.7\% &   3.1\% &   4.6\% &   7.3\% &   5.6\% &   6.5\% &   4.8\% &   5.9\% &   9.5\% &  12.7\% &  11.3\% &  12.0\% &  10.8\% &   1.5\% \\
Bancor                   &    2061 &      52 &     756 &    459 &      6 &      7 &      2 &    242 &     37 &     34 &      2 &      5 &      1 &    166 &     79 &     49 &     13 &     28 &     49 &     14 &      8 &     23 &     18 &     11 &      0 &      0 &      0 &      0 &      0 &      0 &      0 &      0 &      0 &      0 \\
                         &    0.3\% &  100.0\% &  100.0\% &  92.7\% &   2.6\% &   1.9\% &   0.3\% &  27.0\% &   6.3\% &   9.1\% &   1.1\% &   1.7\% &   0.2\% &  26.7\% &   3.7\% &   3.7\% &   1.4\% &   2.7\% &   1.6\% &   0.2\% &   0.1\% &   0.1\% &   0.1\% &   0.0\% &   0.0\% &   0.0\% &   0.0\% &   0.0\% &   0.0\% &   0.0\% &   0.0\% &   0.0\% &   0.0\% &   0.0\% \\
Uniswap V1               &   14304 &       0 &       0 &     36 &    223 &    358 &    743 &    654 &    552 &    341 &    182 &    290 &    478 &    455 &   2038 &   1288 &    949 &   1024 &   3079 &   1121 &    317 &    139 &      1 &     20 &      0 &      1 &      0 &      0 &      4 &      6 &      0 &      1 &      4 &      0 \\
                         &    1.9\% &    0.0\% &    0.0\% &   7.3\% &  97.4\% &  98.1\% &  99.7\% &  73.0\% &  93.7\% &  90.9\% &  98.9\% &  98.3\% &  99.8\% &  73.3\% &  96.3\% &  96.3\% &  98.6\% &  97.3\% &  98.1\% &  18.7\% &   2.5\% &   0.6\% &   0.0\% &   0.0\% &   0.0\% &   0.0\% &   0.0\% &   0.0\% &   0.0\% &   0.0\% &   0.0\% &   0.0\% &   0.0\% &   0.0\% \\
Uniswap V2               &  688466 &       0 &       0 &      0 &      0 &      0 &      0 &      0 &      0 &      0 &      0 &      0 &      0 &      0 &      0 &      0 &      0 &      0 &     10 &   4856 &  12202 &  23231 &  34057 &  54882 &  41559 &  48534 &  34214 &  43105 &  68861 &  91652 &  80297 &  79639 &  62687 &   8680 \\
                         &   91.7\% &    0.0\% &    0.0\% &   0.0\% &   0.0\% &   0.0\% &   0.0\% &   0.0\% &   0.0\% &   0.0\% &   0.0\% &   0.0\% &   0.0\% &   0.0\% &   0.0\% &   0.0\% &   0.0\% &   0.0\% &   0.3\% &  81.1\% &  97.4\% &  99.3\% &  99.3\% &  99.8\% &  99.8\% &  99.6\% &  95.0\% &  96.8\% &  96.7\% &  96.5\% &  94.4\% &  88.3\% &  77.4\% &  76.6\% \\
Sushiswap attacks        &   27243 &       0 &       0 &      0 &      0 &      0 &      0 &      0 &      0 &      0 &      0 &      0 &      0 &      0 &      0 &      0 &      0 &      0 &      0 &      0 &      0 &      0 &    230 &     67 &    100 &    213 &   1782 &   1408 &   2353 &   3298 &   4185 &   5884 &   6549 &   1174 \\
                         &    3.6\% &    0.0\% &    0.0\% &   0.0\% &   0.0\% &   0.0\% &   0.0\% &   0.0\% &   0.0\% &   0.0\% &   0.0\% &   0.0\% &   0.0\% &   0.0\% &   0.0\% &   0.0\% &   0.0\% &   0.0\% &   0.0\% &   0.0\% &   0.0\% &   0.0\% &   0.7\% &   0.1\% &   0.2\% &   0.4\% &   5.0\% &   3.2\% &   3.3\% &   3.5\% &   4.9\% &   6.5\% &   8.1\% &  10.4\% \\
Uniswap V3               &   18455 &       0 &       0 &      0 &      0 &      0 &      0 &      0 &      0 &      0 &      0 &      0 &      0 &      0 &      0 &      0 &      0 &      0 &      0 &      0 &      0 &      0 &      0 &      0 &      0 &      0 &      0 &      0 &      0 &      0 &    613 &   4628 &  11737 &   1477 \\
                         &    2.5\% &    0.0\% &    0.0\% &   0.0\% &   0.0\% &   0.0\% &   0.0\% &   0.0\% &   0.0\% &   0.0\% &   0.0\% &   0.0\% &   0.0\% &   0.0\% &   0.0\% &   0.0\% &   0.0\% &   0.0\% &   0.0\% &   0.0\% &   0.0\% &   0.0\% &   0.0\% &   0.0\% &   0.0\% &   0.0\% &   0.0\% &   0.0\% &   0.0\% &   0.0\% &   0.7\% &   5.1\% &  14.5\% &  13.0\% \\
\bottomrule
\end{tabular}
}
\label{tab:sandwich_stat}
\end{table*}

\subsubsection{Sandwich Gas Prices}
We observe that~\SandwichPercentageOneWei of the back-running transactions ($T_{A2}$) pay only $0$ to $1$ GWei less than $T_{V}$'s gas price (cf.\ Table~\ref{tab:sandwich_a2_gas_price})\footnote{Note that we only consider the~$\numprint{510476}$ sandwich attacks with positive adversarial gas price.}. Intuitively, the closer $T_{A2}$ and $T_{V}$ are, the higher the attacks' success rate due to a chance of other transaction interference. For the front-running transaction ($T_{A1}$), the adversary must also consider the competing sandwich attacker. Given a multi-adversary game, Daian \etal~\cite{daian2020flash} have outlined two primary gas-bidding adversarial strategies: \emph{reactive counter-bidding} and \emph{blind raising}. Under reactive counter-bidding, an adversary only increases its gas price when another competing transaction pays a higher gas price. In blind raising, the adversary raises the gas price of its transaction in anticipation of a raise of its competitors, without necessarily observing competing transactions yet. Recall that geth only accepts an increase of the gas price by at least $10\%$.

When assuming that all attackers adopt the reactive counter-bidding strategy, based on the past sandwich attacks, we estimate that at least~\SandwichPercentageBumpFive of the sandwiches went through more than five rounds of bidding (cf.\ Table~\ref{tab:sandwich_a1_gas_price}). This is because the first $T_{A1}$ bid only needs to add~$1$ Wei to $T_{V}$'s gas price, then each subsequential bid must raise the gas price by $10\%$. After five rounds of bidding, the adversary needs to pay a gas price of at least~$(110\%)^4 \times (GasPrice_V + 1)$ Wei. Fig.~\ref{fig:sandwich-attackers} visualizes the number of adversarial sandwich attack smart contracts we detected. In particular, from the 10th to the 11th of August~2020 (Block~10630000-10640000), we identified~$49$ smart contract addresses attempting to extract value simultaneously.


\begin{table}[tb]
\centering
\caption{The gas price paid by the adversaries for the front-running sandwich transaction $T_{A1}$. A previous study suggests that $79\%$ of the miners (using geth) configure a price bump percentage of $10\%$ to replace an existing transaction from the mempool, while $16\%$ of the miners (using parity) set $12.5\%$ as replacement threshold~\cite{zhou2021high}. Assuming a price bump percentage of $10\%$, we estimate that at least~\SandwichPercentageBumpFive of the attacks experienced more than $5$ counter-reactive bids~\cite{daian2020flash}.}
\resizebox{0.8\columnwidth}{!}{%
\begin{tabular}{lrrc}
\toprule
$r = \frac{GasPrice_{T_{A1}}}{GasPrice_{T_{V}}}$ & \textbf{Count} & \textbf{Percentage} & \textbf{Estimated Bids} \\
\midrule
                                      $r \leq 1$ &     $\numprint{80392}$ &     $15.75\%$ &              $1$ \\
                                $1 < r \leq 1.1$ &  $\numprint{265330}$ &    $51.98\%$ &              $1$ \\
                            $1.1 < r \leq 1.1^2$ &   $\numprint{39963}$ &    $7.83\%$ &              $2$ \\
                          $1.1^2 < r\leq 1.1^3$ &   $\numprint{17014}$ &     $3.33\%$ &              $3$ \\
                          $1.1^3 < r \leq 1.1^4$ &    $\numprint{10223}$ &     $2.00\%$ &              $4$ \\
                                      $1.1^4<r$ &   $\numprint{97554}$ &    $19.11\%$ &            $>=5$ \\
                                          total & $\numprint{510476}$  &   $100.00\%$ &           None \\
\bottomrule
\end{tabular}%
}
\label{tab:sandwich_a1_gas_price}
\end{table}

\begin{table}[tb]
\centering
\caption{Adversarial gas prices for the back-running sandwich transaction $T_{A2}$. \SandwichPercentageOneWei of the transactions pay only $0$ to $1$ GWei less than $T_V$.}
\resizebox{0.8\columnwidth}{!}{%
\begin{tabular}{lrr}
\toprule
    $d = GasPrice_{T_V}-GasPrice_{T_{A2}}$ &  \textbf{Count} & \textbf{Percentage} \\
\midrule
                      $d<0 \text{ GWei}$ &     $\numprint{14162}$ &     $2.77\%$ \\
  $0 \text{ GWei}\leq d <1 \text{ GWei}$ &  $\numprint{408522}$ &    $80.03\%$ \\
  $1 \text{ GWei}\leq d <10 \text{ GWei}$ &    $\numprint{6813}$ &     $1.33\%$ \\
$10 \text{ GWei}\leq d <100 \text{ GWei}$ &   $\numprint{43194}$ &     $8.46\%$ \\
                 $100 \text{ GWei}\leq d$ &    $\numprint{37785}$ &     $7.40\%$ \\
                                    Total &  $\numprint{510476}$ &   $100.00\%$ \\
\bottomrule
\end{tabular}%
}
\label{tab:sandwich_a2_gas_price}
\end{table}

\subsection{Clogging}\label{app:clogging}
Eskandir \etal~\cite{eskandari2019sok} have observed smart contract games which follow the \emph{The War of Attrition}~\cite{fomo3dwiki, ethex:online}. In such a game, players can bid into a pool of money. Each bid resets a timeout, which, once expired, grants the last bidder the entirety of the amassed money. Economists and evolutionary biologists have studied such games for decades~\cite{shubik1971dollar}, and shown that humans overbid significantly. To participate in such contests, users are likely to construct dedicated bidding bots. Those bots are then configured with a specific budget to pay for transaction fees. If an adversary manages to clog the blockchain, such that those bots run out of funding, the attacker can win the bidding game. This is what appears to have happened with the infamous Fomo3D game, where an adversary realized a profit of $\numprint{10469}$ ETH by conducting a clogging attack over~$66$ consecutive blocks (from block \block{6191962} to \block{6191896}).

The throughput of permissionless blockchains is typically limited to about~$7$-$14$ transactions per second, and transaction fee bidding contests have shown to raise the average transaction fees well above~$50$ USD. A \emph{clogging attack} is, therefore, a malicious attempt to consume block space to prevent the timely inclusion of other transactions. To perform a clogging attack, the adversary needs to find an opportunity (e.g., a liquidation, gambling, etc.) which does not immediately allow to extract monetary value. The adversary then broadcasts transactions with high fees and computational usage to congest the pending transaction queue. Clogging attacks on Ethereum can be successful because $79$\% of the miners order transactions according to the gas price~\cite{zhou2021high}.

\subsubsection{Heuristics} To identify past clogging period, we apply the following heuristics.
\begin{itemize}[left=0pt]
\item \textbf{Heuristic 1:} The same address (user/smart contract) consumes more than $80\%$ of the available gas in every block during the clogging period.
\item \textbf{Heuristic 2:} The clogging period lasts for at least five consecutive blocks. Empirical data suggests that the average block time is $13.5\pm0.12$~seconds \cite{ethereum-block-time}, a clogging period of five blocks, therefore, lasts around $1$~minute.
\end{itemize}


\subsubsection{Empirical Results}
We identify ~\CloggingNumAttacks clogging periods from block~\StartBlock to~\EndBlock, where ~\CloggingNumEOA user addresses and~\CloggingNumSC smart contracts are involved (cf.\ Table~\ref{tab:clogging_stat}). While the longest clogging period lasts for $5$ minutes ($24$ blocks), most of the clogging periods ($83.18\%$) account for less than $2$ minutes ($10$ blocks).

\begin{table}[bt!]
\centering
\caption{Detected clogging periods.}
\resizebox{\columnwidth}{!}{%
\begin{tabular}{lrrr}
\toprule
                  Duration &  Clogging Detected &  Avg. Gas Used & Avg. Cost \\
\midrule
$5\sim 9$ blocks ($1\sim 2$ mins) &                $270$ &       $50413969$ &    8 ETH ($12$K USD)\\
$10\sim 14$ blocks ($2\sim 3$ mins) &                 $38$ &      $137697972$ &    54 ETH ($117$K USD)\\
$15\sim 19$ blocks ($3\sim 4$ mins) &                 $10$ &      $199507881$ &   90 ETH ($188$K USD)\\
$20\sim 24$ blocks ($4\sim 5$ mins) &                  $9$ &      $278092700$ &  143 ETH ($326$K USD)\\
$25\sim 29$ blocks ($5\sim 6$ mins) &                  $3$ &      $348737828$ &  369 ETH ($854$K USD)\\
$30\sim 34$ blocks ($6\sim 7$ mins) &                  $2$ &      $458057340$ &  250 ETH ($551$K USD)\\
$35\sim 39$ blocks ($7\sim 8$ mins) &                  $1$ &      $528647491$ &  $297$ ETH ($739$K USD)\\
\bottomrule
\end{tabular}
}
\label{tab:clogging_stat}
\end{table}

\point{Case Studies}
While our heuristics can successfully detect blockchain clogging, they do explain their motivation and we hence manually inspect $14$ selected clogging events (cf.\ Table~\ref{tab:clogging_case_study}). We find that the top~$7$ longest clogging events are related to the non-fungible tokens (NFT), while it's unclear how the adversaries might profit from these events.

\begin{table}[bt!]
\centering
\caption{Selected clogging events.}
\resizebox{\columnwidth}{!}{%
\begin{tabular}{llccccc}
\toprule
\multicolumn{1}{c}{\textbf{Address}} & \textbf{\begin{tabular}[c]{@{}c@{}}Start\\ Block\end{tabular}} & \multicolumn{1}{c}{\textbf{\begin{tabular}[c]{@{}c@{}}Duration\\ (Blocks)\end{tabular}}} & \multicolumn{1}{c}{\textbf{\begin{tabular}[c]{@{}c@{}}Avg. Gas\\ Consumed\end{tabular}}} & \multicolumn{1}{c}{\textbf{\begin{tabular}[c]{@{}c@{}}Avg. Gas\\ Price\end{tabular}}} & \multicolumn{1}{c}{\textbf{\begin{tabular}[c]{@{}c@{}}Cost\\ (ETH)\end{tabular}}} & \multicolumn{1}{c}{\textbf{Usage}} \\
\midrule
\abbrEtherscanAddress{0x099689220846644F87D1137665CDED7BF3422747} &      12953443 &      37 &                95.38\% &              547 &       297.16 & NFT  \\
\abbrEtherscanAddress{0xD4d871419714B778eBec2E22C7c53572b573706e} &      12910380 &      34 &                94.48\% &              790 &       388.16 & NFT  \\
\abbrEtherscanAddress{0x004f5683e183908D0f6b688239e3e2D5bbb066CA} &      12885177 &      31 &                93.99\% &              252 &       112.37 & NFT  \\
\abbrEtherscanAddress{0x18Df6C571F6fE9283B87f910E41dc5c8b77b7da5} &      12934303 &      26 &                90.35\% &  \numprint{1477} &       517.47 & NFT  \\
\abbrEtherscanAddress{0x3a8778A58993bA4B941f85684D74750043A4bB5f} &      12717845 &      25 &                95.85\% &              364 &       130.61 & NFT  \\
\abbrEtherscanAddress{0x18c7766A10df15Df8c971f6e8c1D2bbA7c7A410b} &      12911156 &      25 &                90.05\% &  \numprint{1358} &       459.93 & NFT  \\
\abbrEtherscanAddress{0x3a8778A58993bA4B941f85684D74750043A4bB5f} &      12717893 &      24 &                92.62\% &              503 &       168.03 & NFT  \\

\abbrEtherscanAddress{0x667061051d70b8B4077adBB107d1d98d05A33A4a} &      7091122 &      24 &                91.22\% &              31 &       5.48 & Incentivised clogging  \\
\abbrEtherscanAddress{0xdAC17F958D2ee523a2206206994597C13D831ec7} &     10130772 &      21 &                96.09\% &              40 &       8.05 & Mass USDT transfers     \\
\abbrEtherscanAddress{0xA86958dbD35Ea4D1b55bbD3788B7Aa20D3780AB1} &      8259506 &      15 &                92.59\% &              26 &       3.14 & ETH CAT Attack         \\
\abbrEtherscanAddress{0x67a6C6F0e5ee3E68484B57A3678D5531606721d2} &      7788021 &      15 &                93.21\% &              32 &       3.72 & ERD (E) Attack         \\
\abbrEtherscanAddress{0xA86958dbD35Ea4D1b55bbD3788B7Aa20D3780AB1} &      8260063 &      14 &                94.48\% &              26 &       2.98 & ETH CAT Attack         \\
\abbrEtherscanAddress{0xdAC17F958D2ee523a2206206994597C13D831ec7} &      8509481 &      11 &                89.27\% &              28 &       2.27 & Mass USDT transfers     \\
\abbrEtherscanAddress{0xA86958dbD35Ea4D1b55bbD3788B7Aa20D3780AB1} &      8260051 &      11 &                97.28\% &              26 &       2.41 & ETH CAT Attack         \\
\bottomrule
\end{tabular}
}
\label{tab:clogging_case_study}
\end{table}

    \point{Incentivised clogging} We detect a gambling contract ``Lucky Star'' clogging, where~$203$ addresses perform~$387$ transactions. This game draws the winners, when the cumulative lottery tickets sold exceeds a pre-configured threshold. For every~$\numprint{30000}$ ETH of lottery tickets sold, the accumulated prize is split among the last~$50$ purchasers, the protocol, therefore, incentivizes its users to congest the network at the fictive deadline.
    \point{Attacks on gambling protocols} We also find four clogging events related to two FoMo3D games, namely ETH CAT (cf.\ \abbrEtherscanTx{0x42cead70158235a6ca4868f3cfaf600c7a7b0ebb}) and ERD (E) (cf.\ \abbrEtherscanTx{0x2c58b11405a6a8154fd3bbc4ccaa43924f2be769}). The rules of these gambling protocols is similar to FoMo3D. If no user address purchases a lottery ticket within a fixed time period, the last participant wins the jackpot. We identify two contracts involved in these four clogging events. To ensure that the winner is not already drawn, both contracts have a function to check the current round's status in the corresponding gambling smart contract before they start to spam transactions. These two contracts are deployed by the same address (cf.\ \abbrEtherscanTx{0xfefe743b43291d7799be6e748e9ff4fb0c72aa5c}).
    \point{Mass USDT transfers} We find that two clogging events perform a large number of USDT transfers, wherein~$\numprint{2462}$/$\numprint{1868}$ Ethereum addresses made~$\numprint{2463}$/$\numprint{2032}$ transactions, consuming~$96.07\%$/$89.27\%$ of the gas respectively. Although these activities appear abnormal, we cannot seem to figure out the reason for such behavior.


\section{Transaction Replay Extensions}
\subsection{Replayable Transactions Case Study}\label{app:replaycasestudy}
In Table~\ref{tab:replayable_transaction_case_study}, we present the top~$15$ replayable transactions that produce more than~$100$~ETH and manually classify their motive. We notice~$3$~replayable transactions associated with a previous DeFi attack, the Eminence exploit~\cite{DeFiDege78:online}. It appears that the attacker(s) did not consider the threat of replay transactions. Except for the Eminence exploit, we notice that the bZx attack~\cite{qin2021attacking} transaction is also replayable. We further find~$3$~replayable transactions that invoke the same \emph{DSSLeverage} smart contract (cf.\ \abbrEtherscanAddress{0x4c14eD0b3caC97939053be31150bDbB6f1dDbCA2}). From the DSSLeverage source code, we find that it allows any address to close the contract's position in MakerDAO and retrieve its balance. This coding pattern matches the \emph{sender benefits} pattern (cf.\ Section~\ref{sec:replay-algorithm}). We also discover one on-chain game transaction (Crypto Fishing~\cite{TheFirst52:online}) and five arbitrage transactions. For three of the top~$15$ replayable transactions, we find that the trader is purchasing ERC20 tokens at a favorable price (i.e., arbitrage), as we convert the gained assets back to ETH for our evaluation. 

\begin{table}[tb!]
    \centering
    \caption{Case studies of the top~$15$ non-reverted replayable transactions that yield a profit of more than~$100$~ETH.}
    \resizebox{0.8\columnwidth}{!}{%
    \begin{tabular}{@{}cccc@{}}
    \toprule
    \multicolumn{1}{c}{\textbf{\begin{tabular}[c]{@{}c@{}}Transaction\\hash\end{tabular}}} & \textbf{\begin{tabular}[c]{@{}c@{}}Profit\\(ETH)\end{tabular}} & \textbf{\begin{tabular}[c]{@{}c@{}}Required upfront\\ capital (ETH)\end{tabular}} & \textbf{Motive} \\ \midrule
    \abbrEtherscanTx{0x045b60411af18114f1986957a41296ba2a97ccff75a9b38af818800ea9da0b2a} & $\numprint{16736.9}$ & $0$ & Eminence exploit~\cite{DeFiDege78:online} \\
    \abbrEtherscanTx{0x3503253131644dd9f52802d071de74e456570374d586ddd640159cf6fb9b8ad8} & $\numprint{16393.3}$ & $0$ & Eminence exploit~\cite{DeFiDege78:online} \\
    \abbrEtherscanTx{0x4f0f495dbcb58b452f268b9149a418524e43b13b55e780673c10b3b755340317} & $\numprint{8555.8}$ & $0$ & Eminence exploit~\cite{DeFiDege78:online} \\
    
    \abbrEtherscanTx{0xa85b5edc72bf9ecc4705dd880282961dd3aa38bb33ca56c6c4fd83ae70659a83} & $448.1$ & $0.036$ & ---  \\ 
    \abbrEtherscanTx{0x148f5b58a721aefa230c60e9b1003cf88673a8848fa484aba1739c4a1594533f} & $224.0$ & $0.036$ & ---  \\ 
    \abbrEtherscanTx{0xbab82aa9b5845f658f46b86a7ed3e50bf316e51fcf38a14f20527b3ad785e372} & $183.6$ & $8.0$ & Arbitrage \\
    
    \abbrEtherscanTx{0x40219fdaff2c1575c5fcf0f764ed437b4c5ab69d9f6f8ca6f181b849838a1f89} & $153.2$ & $2.0$ & Arbitrage \\
    \abbrEtherscanTx{0xe7726863cd71fb97bb37fd0cb04ab0a78217e1cf9b23068a6d1165f38f5fd496} & $153.2$ & $2.0$ & Arbitrage \\
    \abbrEtherscanTx{0x475af146e9d4a0ef111a27aeefb732c14b0934ecbafc0654e02470f6623fcd8f} & $152.5$ & $0$ & DSSLeverage \\
    \abbrEtherscanTx{0xfa5fc8aaa7875a0d606a7c976459de67b75318074275fe6634d2bc6121debb03} & $144.3$ & $0$ & DSSLeverage \\
    \abbrEtherscanTx{0x2e274d302c9bf7f566510f236d74b02aff5cae9ebda5313cf26c0cdc59f8ee45} & $136.3$ & $0$ & DSSLeverage \\
    
    \abbrEtherscanTx{0x7ca2bd7ddd0413a75882a0cfa4c1900afd4061ffd10b611b24013e5d1b990765} & $129.8$ & $9.0$ & Arbitrage \\
    
    \abbrEtherscanTx{0xd46c6657bbffcf09dfdbca5050f0c60c3f7eb92a2c3de8c9e5c7c53425ebb091} & $118.0$ & $5.0$ & Crypto Fishing~\cite{TheFirst52:online} \\
    
    \abbrEtherscanTx{0xc2f3001f87069444c86e2b2567727eb503d7a95d8282289fab1ebefa057acac8} & $112.0$ & $0.036$ & --- \\
    \abbrEtherscanTx{0x9f4b020f44cfcb8d855713d9bff567a522421949ee50b08269fa2e7865baec7e} & $106.3$ & $0.80609$ & Arbitrage \\
    
    
    \bottomrule
    \end{tabular}%
    }
    \label{tab:replayable_transaction_case_study}
\end{table}

\subsection{Replay Protection}\label{app:replay-protection}
Listing~\ref{lst:naive_replay_protections} presents the solidity snippets that mitigates the transaction replay attack (cf.\ Section~\ref{sec:replay-algorithm}).
\begin{lstlisting}[float,floatplacement=H,label=lst:naive_replay_protections,language=Solidity,caption={Protection from the transaction replay attack.}]
pragma solidity ^0.6.0;

contract ReplayProtections {
  address owner;
    
  constructor () {
    owner = 0x00..33;
  }
    
  function Authentication() public {
    require(msg.sender == owner);
    uint profit;
    // profiting logic omitted for brevity
    msg.sender.transfer(profit);
  }
	
  function MoveBeneficiary() public {
    address beneficiary = 0x01..89;
    uint profit;
    // profiting logic omitted for brevity
    beneficiary.transfer(profit);
  }
}
\end{lstlisting}

\section{Privately Relayed Transaction Measurement}\label{app:privatetransactions}

\subsection{Identifying Non-Broadcast Transactions}
\begin{figure}[tb]
    \centering
    \includegraphics[width=\columnwidth]{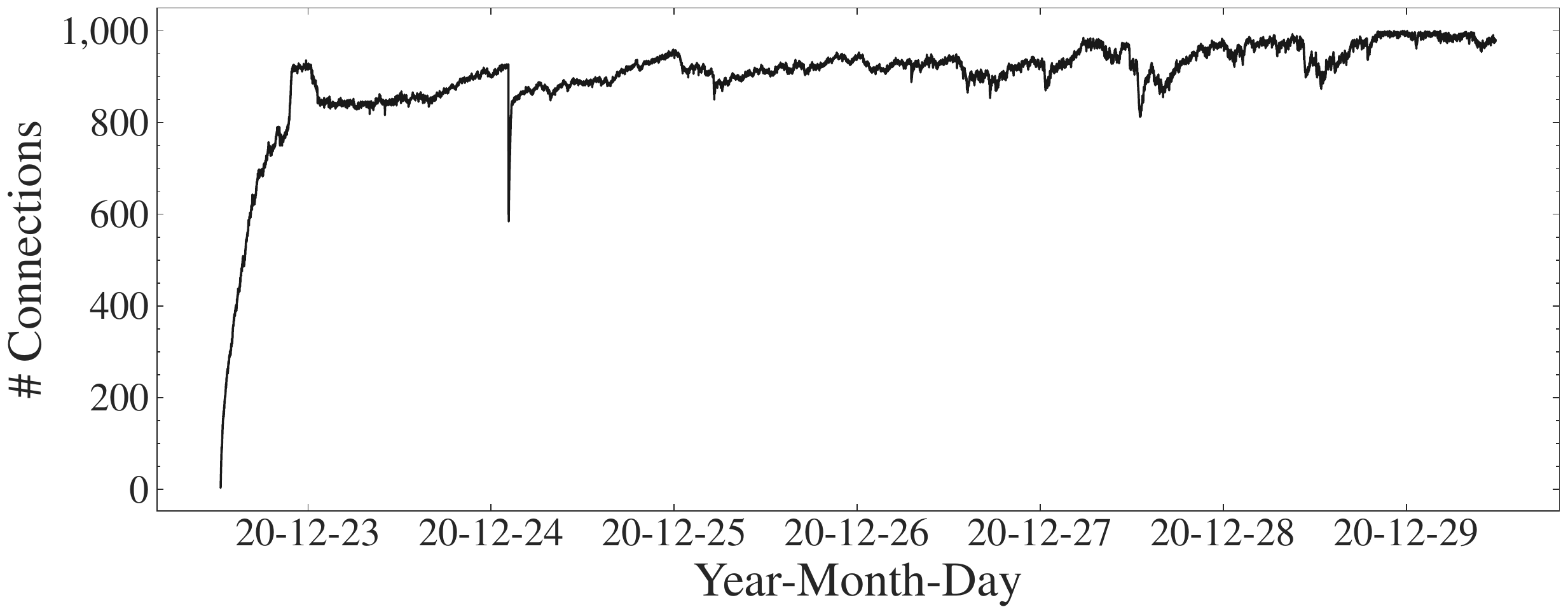}
    \caption{Number of connections of our modified geth node while listening for transactions on the P2P network. The default geth configuration maintains 50 connections. The more connections a node manages, the earlier this node receives block and transactions from neighboring peers.}
    \label{fig:gethconnections}
\end{figure}

To measure the fraction of transactions that are mined, but not broadcast on the P2P network, we set up a well connected geth client with at most~$\numprint{1000}$ connections in the Ethereum network (cf.\ Fig.~\ref{fig:gethconnections})\footnote{A default geth client connects to a maximum of~$50$ peers. We remark that our mass-connection client can cover a wide range of peers. This is because peerings in Ethereum are primarily influenced by the distance of the peer nodes’ ID hashes\cite{kim2018measuring}, rather than the physical location, although location does influence latency.}. The client records any new incoming transaction, before it is added to the memory pool, or written to the blockchain. The number of connections of the Ethereum client are important as in to \emph{(i)} receive data as early as possible~\cite{gervais2015tampering} and \emph{(ii)} to maximize an all encompassing view of the network layer. Once we stored all visible transactions, we compare this network layer dataset with the resulting confirmed blockchain transactions to identify the transactions that were mined, but not broadcast.

\subsection{Empirical Results}
When observing the Ethereum P2P network over $\numprint{45669}$ blocks (1 week) from block \block{11503300} (22nd December,~2020) to \block{11548969} (29th December,~2020), the chain recorded~$\numprint{8285218}$ transactions. When comparing those with the transactions we observed on the network layer, we find that~$\numprint{136143}$ mined transactions were not broadcast prior to being mined. We hence can conclude that $1.64\%$ of the transactions are privately relayed. We manually verify $100$ transactions at random from our dataset with the data provided by Etherscan~\cite{etherscan}, and can confirm that our methodology matches the privately relayed transactions reported. We notice that parts of the detected private transactions are payout transactions from mining pool operators to miners. By excluding the transactions that consume~$\numprint{21000}$ gas, we find~$\numprint{11374}$ ($8.35\%$) private transactions invoking smart contracts (cf.\ Table~\ref{tab:privateminers}).~$\numprint{21000}$~is the minimum gas cost of an Ethereum transaction, i.e., a simple transfer costs~$\numprint{21000}$ gas.

\point{Private 1inch Trades}
By observing privately relayed transactions, we identify with which miners 1inch reached private peering agreements. We for instance found two
privately relayed 1inch transactions (cf.\ \abbrEtherscanTx{0xa0263443c173d6d21bb1da0e931456cdbbc6ee4c0c090b689982ef33d44db15b} and \abbrEtherscanTx{0xaa45cc189f75d44ebb2d8cbf56ad763da49f2adef3d82147772ea91f2e3ec66f}) from the Spark Pool ($23.50$\% hashrate), one (cf.\ \abbrEtherscanTx{0xe4d4013852713d42988e897f4a0f3e01f82f96e5385c46fecab1adf5f9c486b5})
from the Babel Pool ($4.83$\% hashrate) and one (cf.\ \abbrEtherscanTx{0x4340470116020410e7e5bfe5e069f512432dddb323744bedcc2f799b8136aeb5})
from the F2Pool ($9.59$\% hashrate).

\begin{table}[tb]
\centering
\caption{Distribution of the number of privately relayed transactions per miner coinbase address over~$\numprint{45669}$ blocks (1 week). Data measured from the P2P network with a geth client which consistently maintains over~$800$ P2P connections (cf.\ Fig.~\ref{fig:gethconnections}). We measure the hashrate based on the number of blocks found during measurement by the respective miner.
}
\resizebox{\columnwidth}{!}{%
\begin{tabular}{@{}lllll@{}}
\toprule
\textbf{Miner address} & \multicolumn{1}{l}{\textbf{\begin{tabular}[c]{@{}l@{}}Private transactions\\ (contract invoking)\end{tabular}}} & \textbf{Name} & \textbf{Hashrate}\\ \midrule
\etherscanAddress{0xEA674fdDe714fd979de3EdF0F56AA9716B898ec8} & $104,674$ $(7,310)$      & Ethermine        & $20.81\%$  &  \\
\etherscanAddress{0x829BD824B016326A401d083B33D092293333A830} & $19,560$ $(329)$       & F2Pool           & $9.59\%$         &  \\
\etherscanAddress{0x99C85bb64564D9eF9A99621301f22C9993Cb89E3} & $5,926$ $(19)$        & BeePool          & $2.11\%$         &  \\
\etherscanAddress{0x5A0b54D5dc17e0AadC383d2db43B0a0D3E029c4c} & $3,256$ $(2,775)$        & Spark Pool       & $23.50\%$         &  \\
\etherscanAddress{0xB3b7874F13387D44a3398D298B075B7A3505D8d4} & $980$ $(568)$         & Babel Pool       & $4.83\%$         &  \\
\etherscanAddress{0xD224cA0c819e8E97ba0136B3b95ceFf503B79f53} & $697$ $(191)$         & UUPool           & $3.46\%$         &  \\
\etherscanAddress{0x5921c6a53c2cD0987Ae111b59F2E5dDaAf275b60} & $360$ $(0)$         & -                & $0.45\%$         &  \\
\etherscanAddress{0x04668Ec2f57cC15c381b461B9fEDaB5D451c8F7F} & $303$ $(1)$         & zhizhu.top/SpiderPool     & $7.76\%$         &  \\
\etherscanAddress{0x314653F5933FC25D0A428424f5A645B2bcc37483} & $142$ $(135)$         & -                & $0.11\%$         &  \\
\etherscanAddress{0x3EcEf08D0e2DaD803847E052249bb4F8bFf2D5bB} & $59$ $(5)$          & MiningPoolHub    & $1.75\%$        &  \\
\etherscanAddress{0x52f13E25754D822A3550D0B68FDefe9304D27ae8} & $59$ $(1)$          & EthashPool 2     & $0.1\%$         &  \\
\etherscanAddress{0xAEe98861388af1D6323B95F78ADF3DDA102a276C} & $58$ $(2)$          & -                & $0.21\%$        &  \\
\etherscanAddress{0x00192Fb10dF37c9FB26829eb2CC623cd1BF599E8} & $25$ $(22)$          & 2Miners: PPLNS   & $2.01\%$        &  \\
\etherscanAddress{0xB35c1055aAE02DA8497E9Dd866e27C86be16CFEF} & $22$ $(0)$          & -                & $0.06\%$         &  \\
\etherscanAddress{0x002e08000acbbaE2155Fab7AC01929564949070d} & $7$ $(7)$           & Hiveon Pool      & $0.95\%$         &  \\
\etherscanAddress{0x1aD91ee08f21bE3dE0BA2ba6918E714dA6B45836} & $7$ $(1)$           & 2Miners: SOLO    & $4.01\%$         &  \\
\etherscanAddress{0x35F61DFB08ada13eBA64Bf156B80Df3D5B3a738d} & $4$ $(4)$           & firepool         & $0.62\%$         &  \\
\etherscanAddress{0x45a36a8e118C37e4c47eF4Ab827A7C9e579E11E2} & $1$ $(1)$           & -                & $0.11\%$         &  \\
\etherscanAddress{0x8595Dd9e0438640b5E1254f9DF579aC12a86865F} & $1$ $(1)$           & EzilPool 2       & $0.68\%$         &  \\
\etherscanAddress{0xF541C3CD1D2df407fB9Bb52b3489Fc2aaeEDd97E} & $1$ $(1)$           & -                & $0.32\%$         &  \\
\etherscanAddress{0x2A0eEe948fBe9bd4B661AdEDba57425f753EA0f6} & $1$ $(1)$           & -                & $0.56\%$         &  \\\midrule
\textbf{Total} & $136,143$ $(11,374)$ & - & $84.00\%$
\\\bottomrule
\end{tabular}
}
\label{tab:privateminers}
\end{table}

\point{Mining Pools Engaging in Private Transactions}
In Table~\ref{tab:privateminers} we provide the distribution of miners engaging in mining non-broadcast transactions. Over the course of $45,669$ blocks ($1$ week), we identified $81$ miners, of which $21$ ($26$\%) mine transactions privately. We notice that the number of privately relayed transactions does not necessarily correspond to the hashing power of the miner. The \emph{Ethermine} miner positions private transactions (e.g., benign mining payouts) at the block start with apparent low gas prices. The \emph{SparkPool}, however, seemingly trying to disguise its private transactions as ordinary instances by paying regular gas prices. We identified for example the following transaction hashes: \abbrEtherscanTx{0x4e173c71d481a94169839a6a0e6b912c2631589db1a7a42596649a692f3a29cd}, \abbrEtherscanTx{0xa67e709687dc64a543387f7219aadc0e7f29f207d838caf2d99fd69b4d684725}. In particular, we noticed the contract \abbrEtherscanAddress{0x000000000025d4386f7fb58984cbe110aee3a4c4}, for which all interacting transactions are mined by the SparkPool and not broadcast on the P2P network. Based on the available EVM byte code and engaging transactions, this contract appears to be involved in trading, strongly indicating that the SparkPool is engaging in MEV before the emergency of BEV relayers.

\point{Private Value Extracting Transactions} From block \block{11503300} to \block{11548969}, we discover~$340$ liquidation transactions on Aave, Compound and dYdX (cf.\ Section~\ref{sec:fixed-spread-liquidation}) out of which we identify~$18$ private transactions. We also detect~$5$ private transactions among the~$\numprint{1067}$ arbitrage transactions.
\point{Private Replayable Transactions}
We find that~$\numprint{1156}$ of the~$\numprint{8285218}$ transactions are replayable following the methodology of Section~\ref{sec:replay-evaluation}. Out of these replayable transactions, we identify~$13$ private transactions yielding a profit of~$0.59$~ETH. Through manually inspection, we find that these~$13$ transactions are 1inch exchange trades. We recall that private transactions cannot be replayed by non-miners.

\end{document}